\documentclass{aa}

\usepackage{graphicx}
\usepackage[varg]{txfonts}
\usepackage{natbib}
\bibpunct{(}{)}{;}{a}{}{,}
\usepackage{array}
\usepackage{xspace}
\usepackage{lscape}
\usepackage{url}
\usepackage{arydshln}
\usepackage[hyperfootnotes=false, linktocpage=true, breaklinks=true, colorlinks=true, linkcolor=blue, citecolor=blue, urlcolor=blue]{hyperref}
\usepackage[all]{hypcap}
\usepackage{longtable}
\usepackage{afterpage}

\newcommand{\kms}{\ensuremath{\mathrm{km\ s^{-1}}}\xspace}
\newcommand{\NH}{\ensuremath{N_{\mathrm{H}}}\xspace}

\newcommand{\xabs}{\xspace{\tt xabs}\xspace}
\newcommand{\xmm}{{\it XMM-Newton}\xspace}
\newcommand{\chandra}{{\it Chandra}\xspace}

\newcommand{\astroh}{{\it Astro-H}\xspace}

\newcommand{\ngc}{{NGC~5548}\xspace}

\newcommand{\hitomi}{{\it Hitomi}\xspace}
\newcommand{\athena}{{\it Athena}\xspace}

\newcommand{\xstar}{{\tt XSTAR}\xspace}
\newcommand{\cloudy}{{\tt Cloudy}\xspace}
\newcommand{\spex}{\xspace{\tt SPEX}\xspace}
\newcommand{\pion}{\xspace{\tt pion}\xspace}

\newcommand{\TC}{\xspace{$T_{\rm C}$}\xspace}

\mathchardef\mhyphen="2D

\begin{document}

\title{Systematic comparison of photoionised plasma codes with application to spectroscopic studies of AGN in X-rays}

\author{
M. Mehdipour \inst{1}
\and
J.S. Kaastra \inst{1,2,3}
\and 
T. Kallman \inst{4}
}

\institute{
SRON Netherlands Institute for Space Research, Sorbonnelaan 2, 3584 CA Utrecht, the Netherlands\\ \email{M.Mehdipour@sron.nl}
\and
Department of Physics and Astronomy, Universiteit Utrecht, P.O. Box 80000, 3508 TA Utrecht, the Netherlands
\and
Leiden Observatory, Leiden University, PO Box 9513, 2300 RA Leiden, the Netherlands
\and
NASA Goddard Space Flight Center, Code 662, Greenbelt, MD 20771, USA
}

\date{Received 15 April 2016 / Accepted 7 October 2016}

\abstract
{
Atomic data and plasma models play a crucial role in the diagnosis and interpretation of astrophysical spectra, thus influencing our understanding of the universe. In this investigation we present a systematic comparison of the leading photoionisation codes to determine how much their intrinsic differences impact X-ray spectroscopic studies of hot plasmas in photoionisation equilibrium. We carry out our computations using the \cloudy, \spex, and \xstar photoionisation codes, and compare their derived thermal and ionisation states for various ionising spectral energy distributions. We examine the resulting absorption-line spectra from these codes for the case of ionised outflows in active galactic nuclei. By comparing the ionic abundances as a function of ionisation parameter $\xi$, we find that on average there is about 30\% deviation between the codes in $\xi$ where ionic abundances peak. For H-like to B-like sequence ions alone, this deviation in $\xi$ is smaller at about 10\% on average. The comparison of the absorption-line spectra in the X-ray band shows that there is on average about 30\% deviation between the codes in the optical depth of the lines produced at ${\log \xi \sim 1}$ to 2, reducing to about 20\% deviation at ${\log \xi \sim 3}$. We also simulate spectra of the ionised outflows with the current and upcoming high-resolution X-ray spectrometers, on board \xmm, \chandra, \hitomi, and \athena. From these simulations we obtain the deviation on the best-fit model parameters, arising from the use of different photoionisation codes, which is about 10 to 40\%. We compare the modelling uncertainties with the observational uncertainties from the simulations. The results highlight the importance of continuous development and enhancement of photoionisation codes for the upcoming era of X-ray astronomy with \athena.
}

\authorrunning{M. Mehdipour et al.}
\titlerunning{Systematic comparison of photoionisation codes}
\maketitle

\section{Introduction}

An astrophysical object with an intense continuum radiation strongly influences the ionisation and thermal state of its nearby gas. For example, this is the case in active galactic nuclei (AGN), where accretion of matter onto a supermassive black hole (SMBH) releases a huge amount of radiation, leading to the photoionisation of the surrounding gas outflows. Such a medium is generally treated as in photoionisation equilibrium (PIE), and thus the ionisation state of the plasma is primarily regulated by the balance between photoionisation and recombination. Photoionised plasmas are however complex environments to model because of various processes that play a role in reaching photoionisation equilibrium. The equilibrium electron temperature $T$ of a PIE plasma is determined by the solution to the energy balance equation, where the rate of energy injection into the plasma (e.g. photoelectrons) is set equal to the rate of energy loss from the plasma (e.g. radiation by radiative recombination).

The ionisation parameter $\xi$ \citep{Tar69, Kro81} conveniently quantifies the ionisation state of a PIE plasma with a single parameter, which is defined as 
\begin{equation}
\label{xi_eq}
\xi \equiv \frac{L}{{n_{\rm{H}}\,r^2 }}
,\end{equation}
where $L$ is the luminosity of the ionising source over the 1--1000 Ryd (13.6 eV to 13.6 keV) band in $\rm{erg}\ \rm{s}^{-1}$, $n_{\rm{H}}$ the hydrogen density in $\rm{cm}^{-3}$, and $r$ the distance between the plasma and ionising source in cm.

Given the definition of $\xi$, a self-consistent solution to the ionisation and energy balance equations yields the temperature and ionic abundances of a PIE plasma as a function $\xi$. The photoionisation codes, namely \cloudy\footnote{\url{http://www.nublado.org}} \citep{Fer13}, \spex\footnote{\url{http://www.sron.nl/spex}} \citep{Kaa96}, and \xstar\footnote{\url{http://heasarc.gsfc.nasa.gov/xstar/xstar.html}} \citep{Kall01, Bau01}, compute this thermal and ionisation balance based on the spectral energy distribution (SED) of the ionising source and the elemental abundances of the ionised plasma. The codes take the vast database of atomic data into account to derive the solution between various heating and cooling mechanisms, such as photoionisation, recombination, Auger ionisation, collisional ionisation, bremsstrahlung, and Compton scattering. For a classical paper describing a PIE plasma and modelling its relevant processes, see \citet{Kall82}. For a review of atomic data used in the modelling of hot plasmas, see \citet{Kall07}.

In this investigation we used \cloudy version 13.01, \spex version 3.02.00, and \xstar version 2.3 to carry out a systematic comparison of the results from these photoionisation codes. In Sect. \ref{pie_sect} we describe the PIE calculations via the three codes for different SEDs. In Sect. \ref{s_curve_sect} we determine the thermal state of PIE plasmas using each code and present their thermal stability analysis. In Sect. \ref{heat_cool_sect} we study how different processes contribute to the cooling and heating of a PIE plasma, and how they change for different SED cases. In Sect. \ref{ion_frac_sect} we compare the ionisation state of PIE plasmas computed by the three codes. We analyse the corresponding transmission spectra in Sect. \ref{spectra_sect}, and determine the spectral-line differences found by the three codes. In Sect. \ref{obs_sect}, for the study of ionised outflows in AGN, we compare the uncertainties arising from modelling with different photoionisation codes with the statistical uncertainties from observations with high-resolution X-ray spectrometers. We discuss all our findings in Sect. \ref{discussion} and give concluding remarks in Sect. \ref{conclusions}. In Appendix \ref{appendix} we provide a table, comparing the ionic abundances found by the three codes.

\section{Photoionisation equilibrium calculations}
\label{pie_sect}

We computed the ionisation balance in \cloudy, \spex, and \xstar using the four SEDs described in Sect. \ref{SED_sect} and the elemental abundances described in Sect. \ref{abund_sect}. In \spex, we used the new \pion model for photoionisation calculations, which is introduced in Sect. \ref{pion_sect}. In our ionisation balance calculations with each code, we adopt an optically thin photoionised plasma in equilibrium with a slab geometry. The hydrogen density was set to ${n_{\rm{H}} = 1 \times 10^{8}}$~${\rm{cm}}^{-3}$, with a total column density of ${N_{\rm{H}} = 1 \times 10^{16}}$~${\rm{cm}}^{-2}$. Later in Sects. \ref{spectra_sect} and \ref{obs_sect}, where we calculate the absorption-line spectra of photoionised plasma, the column density is set to ${N_{\rm{H}} = 1 \times 10^{22}}$~${\rm{cm}}^{-2}$.

We note that up to and including version 13.03 of \cloudy, $L$ in the definition of $\xi$ was taken to be the total ionising luminosity as first defined by \citet{Tar69}. However, from version 13.04 of \cloudy, the default $L$ is changed to be consistent with the commonly used definition, where it ranges between 1 and 1000 Ryd. In this paper, $L$ is always over 1--1000 Ryd in our calculations with each code.

\subsection{Spectral energy distributions}
\label{SED_sect}
%

%
\begin{figure}[!tbp]
\centering
\resizebox{1.04\hsize}{!}{\hspace{-0.7cm}\includegraphics[angle=0]{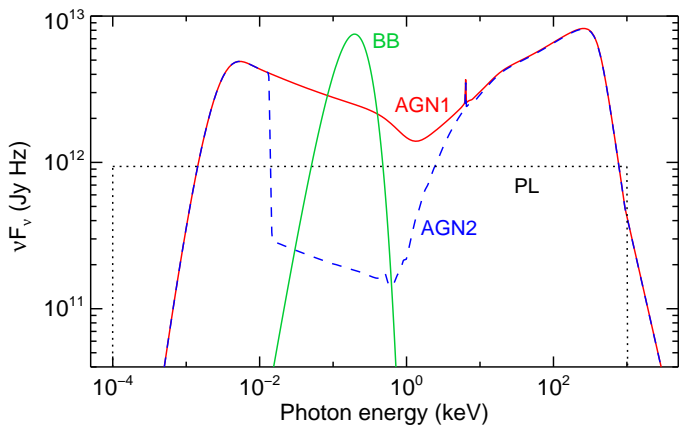}}
\caption{Four different SEDs that we used for our photoionisation balance calculations in \cloudy, \spex, and \xstar. The AGN1 and AGN2 SEDs are the normal and obscured SED versions of a typical Seyfert AGN (\ngc), taken from \citet{Meh15a}. The PL SED is a power-law continuum with ${\Gamma = 2}$, and the BB SED is a black body with ${T = 50}$~eV.}
\label{SED_fig}
\end{figure}

In this paper, we used four different SEDs for our ionisation balance calculations, which are displayed in Fig. \ref{SED_fig}. This enables us to investigate the effects of the ionising SED on the derived results from each code.

The first SED, labelled AGN1, corresponds to that of Seyfert~1 galaxy \object{NGC~5548}, derived in \citet{Meh15a} from modelling extensive multiwavelength campaign data of this object. The AGN1 SED represents the broadband continuum of a standard unobscured AGN. The second SED, labelled AGN2, is the obscured version of AGN1. This SED is also taken from \citet{Meh15a} and represents the broadband continuum after absorption by cold gas at the core of this AGN. The extreme ultraviolet (EUV) and soft X-ray parts of this SED are suppressed as shown in Fig. \ref{SED_fig}. This obscured SED (AGN2) ionises those outflows, which are located further out from the nucleus.

The third SED, labelled PL, corresponds to a simple power-law continuum with ${\Gamma = 2}$, spanning from 0.1 eV to 1 MeV. A power-law SED is sometimes used as an approximation for the SED of those objects, which their broadband continuum model is not established. The fourth SED, labelled BB, corresponds to the spectrum of a simple blackbody emitter with a temperature of ${T = 50}$~eV. This SED is chosen to be in contrast with other SEDs to represent a very soft spectrum. 

\subsection{Elemental abundances}
\label{abund_sect}
For the elemental abundances of the PIE plasma, the proto-solar values of \citet{Lod09} were adopted. The absolute abundances used in our calculations with each code are given in Table \ref{abundance_table}.

%
\begin{table}[!tbp]
\begin{minipage}[t]{\hsize}
\setlength{\extrarowheight}{3pt}
\caption{Absolute abundances of chemical elements that we used in our computations of the photoionisation equilibrium and the transmission spectrum. The values are from proto-solar abundances of \citet{Lod09}.}
\label{abundance_table}
\centering
\small
\renewcommand{\footnoterule}{}
\begin{tabular}{| l | l || l | l |}
\hline \hline
Element & Abundance & Element & Abundance \\
\hline

H       &       $1$                                                      &       S        &      ${1.622 \times 10^{-5}}$        \\
He       &      ${9.705 \times 10^{-2}}$        &   Cl   &      ${1.991 \times 10^{-7}}$       \\
Li       &      ${2.143 \times 10^{-9}}$        &    Ar  &      ${3.573 \times 10^{-6}}$       \\
Be      &       ${2.360 \times 10^{-11}}$       &    K   &      ${1.449 \times 10^{-7}}$       \\
B        &      ${7.244 \times 10^{-10}}$       &    Ca  &      ${2.328 \times 10^{-6}}$       \\
C        &      ${2.773 \times 10^{-4}}$        &    Sc  &      ${1.327 \times 10^{-9}}$       \\
N        &      ${8.166 \times 10^{-5}}$        &    Ti  &      ${9.528 \times 10^{-8}}$       \\
O        &      ${6.053 \times 10^{-4}}$        &    V   &      ${1.102 \times 10^{-8}}$       \\
F        &      ${3.097 \times 10^{-8}}$        &    Cr  &      ${5.047 \times 10^{-7}}$       \\
Ne       &      ${1.268 \times 10^{-4}}$        &    Mn  &      ${3.556 \times 10^{-7}}$       \\
Na       &      ${2.223 \times 10^{-6}}$        &    Fe  &      ${3.266 \times 10^{-5}}$       \\
Mg       &      ${3.972 \times 10^{-5}}$        &    Co  &      ${9.057 \times 10^{-8}}$       \\
Al       &      ${3.258 \times 10^{-6}}$        &    Ni  &      ${1.888 \times 10^{-6}}$       \\
Si       &      ${3.855 \times 10^{-5}}$        &    Cu  &      ${2.084 \times 10^{-8}}$       \\
P        &      ${3.199 \times 10^{-7}}$        &   Zn   &      ${5.012 \times 10^{-8}}$       \\

\hline

\end{tabular}
\end{minipage}

\end{table}

\subsection{The new {\tt pion} model in {\tt SPEX}}
\label{pion_sect}

Here we introduce the new \pion model in \spex, which is a self-consistent photoionisation model that calculates both the ionisation balance and the spectrum. Previously, we used the \xabs model in \spex, which calculated the transmission through a slab of photoionised plasma with all ionic abundances linked in a physically consistent fashion through precalculated runs with external codes (\cloudy or \xstar). However, the new \pion model is developed to calculate all the steps in \spex.

The \pion model uses the ionising radiation from the continuum components set by the user in \spex. So during spectral fitting, as the continuum varies, the ionisation balance and the spectrum of the PIE plasma are recalculated at each stage. This means while using realistic broadband continuum components to fit the data (e.g. Comptonisation and reflection models for AGN), the photoionisation balance and the spectrum are calculated accordingly by the \pion model. Thus, the variable nature of the source continuum can also be taken into account in ionisation balance calculations. So rather than assuming an SED shape for ionisation balance calculations, the \pion model provides a more accurate approach for determining the intrinsic continuum and the ionisation balance. For example, \citet{Chak12} have shown that just the temperature of the accretion disk and the strength of the soft X-ray excess component in AGN (e.g. \citealt{Meh11}) can significantly influence the structure and stability of the ionised outflows in AGN. The \pion model was first used in a recent paper by \citet{Mill15} to model the complex absorption spectrum of ionised flows caused by the tidal disruption of a star by a massive black hole. For a description of the atomic database used in the \pion model, see the \spex manual. 

\section{Thermal state of photoionised plasmas}
\label{s_curve_sect}

\begin{figure}[!tbp]
\centering
\resizebox{1.0\hsize}{!}{\hspace{0cm}\includegraphics[angle=0]{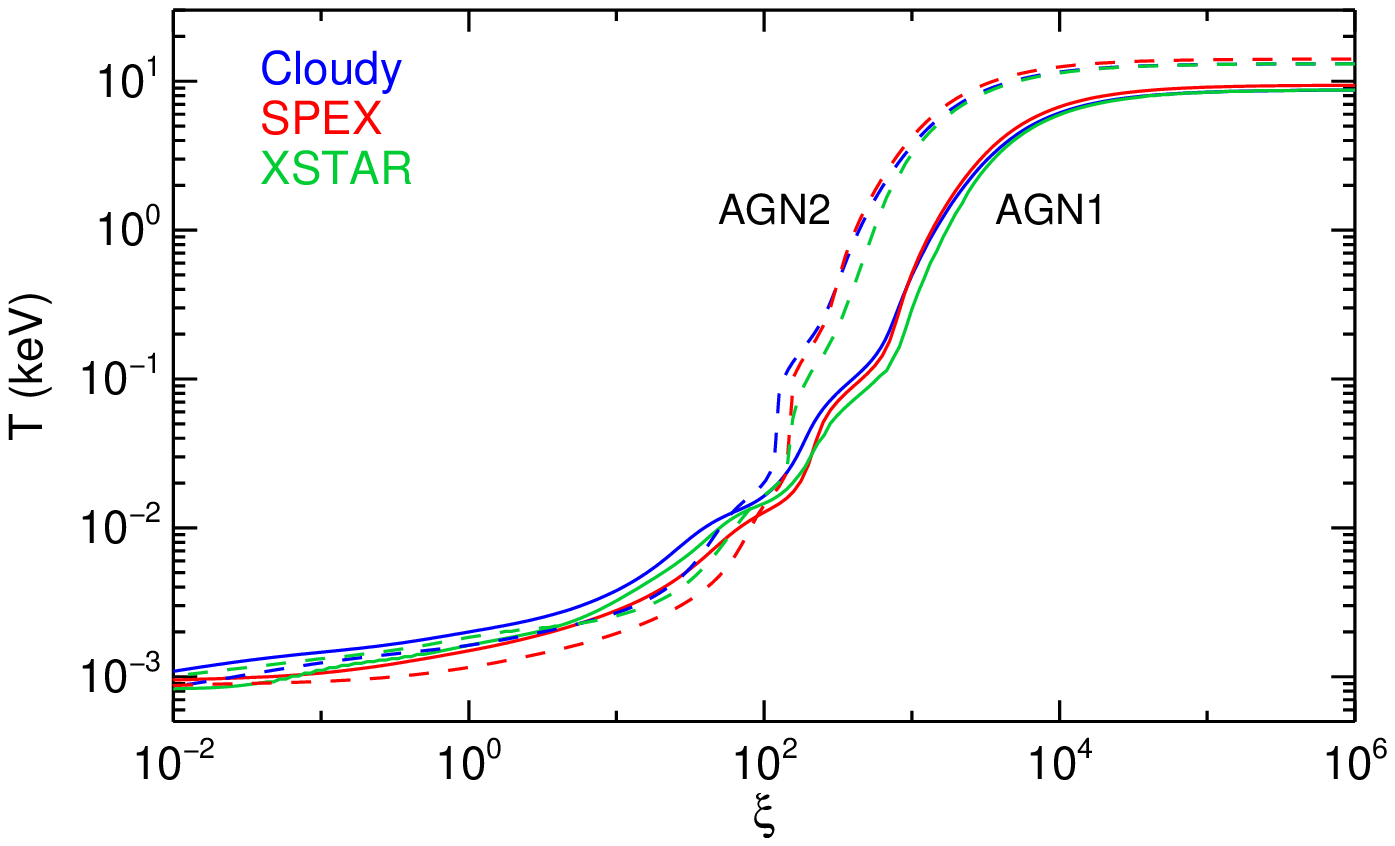}}\vspace{-0.7cm}
\resizebox{1.0\hsize}{!}{\hspace{0cm}\includegraphics[angle=0]{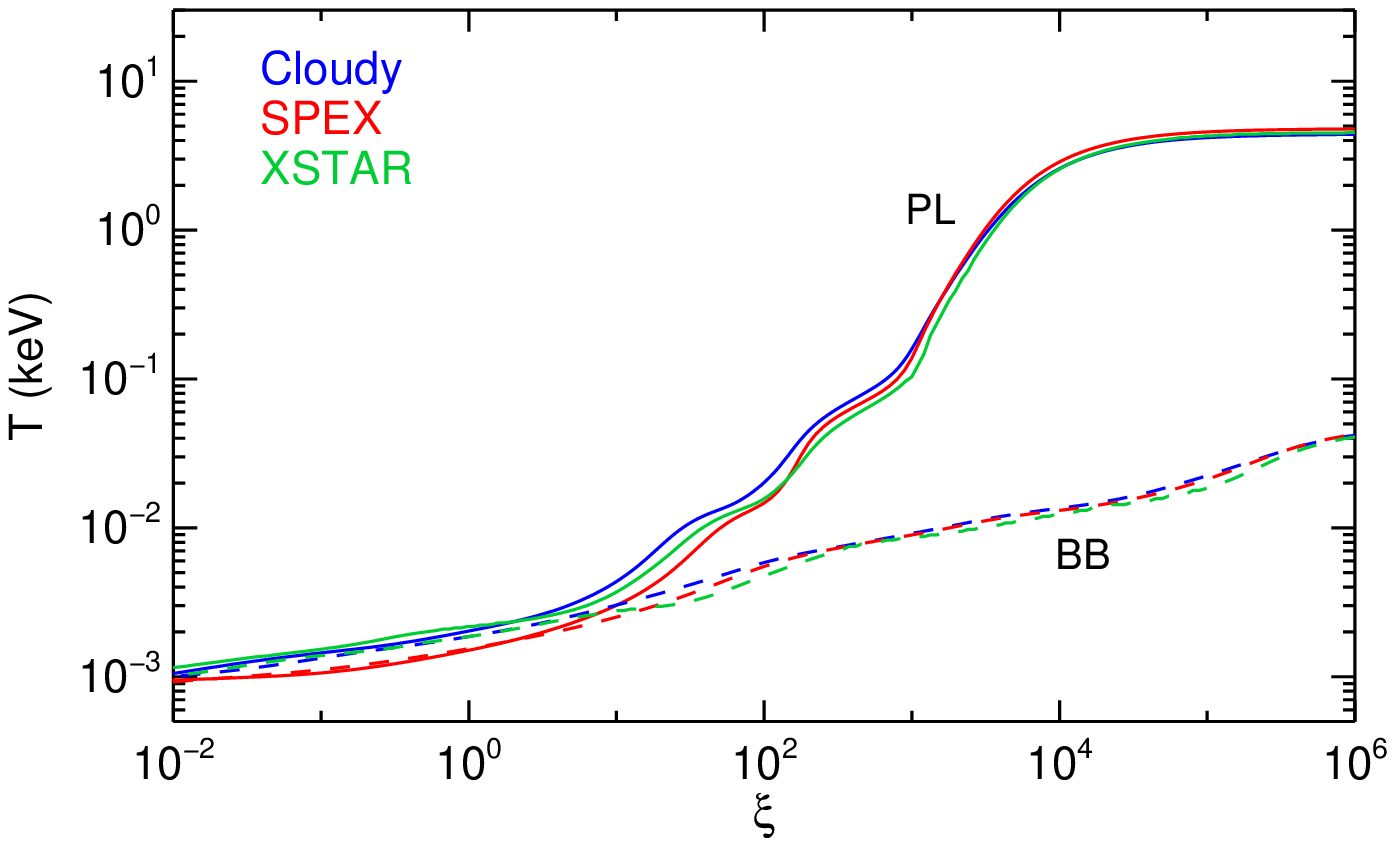}}
\caption{Electron temperature $T$ of a PIE plasma as a function of ionisation parameter $\xi$. The curves are calculated using the \cloudy, \spex, and \xstar photoionisation codes, shown in blue, red, and green, respectively. The calculations are carried out for the four different SEDs of Fig. \ref{SED_fig}: AGN1 (top panel in solid line), AGN2 (top panel in dashed line), PL (bottom panel in solid line), and BB (bottom panel in dashed line).}
\label{T_xi}
\end{figure}

Following the ionisation balance calculations described in Sect. \ref{pie_sect}, here we present the solutions obtained by \cloudy, \spex, and \xstar for the temperature and ionisation of the PIE plasma. In Fig. \ref{T_xi} we show the electron temperature $T$ of the plasma as a function of $\xi$ found by the codes for each of the four SEDs.

Figure \ref{T_xi} demonstrates the different impact of each SED on the ionisation balance of the plasma, and hence $T(\xi)$. The results show that there is reasonable agreement between $T(\xi)$ from the codes for each SED case. We compared the temperatures at $\log \xi$ of 1.0, 2.0, and 3.0, which covers a range commonly found in AGN ionised outflows from X-ray spectroscopy. Over this $\xi$ range, for the AGN1 SED there is between 25\% and 29\% deviation between the maximum and minimum $T$ values found by the codes. For the AGN2 SED, the difference is greater at between 37\% to 45\%, while for the PL SED, it is between 27\% and 36\%. For the BB SED, there is the least amount of $T$ difference at 17\% to 19\%.

There is a reasonably good agreement between the Compton temperatures \TC found by the codes for all four SED cases. The obtained \TC values are given in Table \ref{T_C_table}. The deviation between the codes from the mean \TC is about 4\%. The \TC value is highest for the AGN2 SED case and lowest for the BB case. 

An ionising SED determines the ionisation balance and thermal stability of photoionised plasmas, such as the ionised outflows in AGN. A photoionised plasma can be thermally unstable in certain regions of the ionisation parameter space. This can be investigated by means of producing the thermal stability curves (also called S-curves or cooling curves), which is a plot of the electron temperature $T$ as a function of the pressure form of the ionisation parameter, $\Xi$, introduced by \citet{Kro81}. The ionisation parameter $\Xi$, is defined as $\Xi  \equiv {F}/{{n_{\rm{H}}\, c\, kT}}$, where $F$ is the flux of the ionising source between 1--1000 Ryd (in $\rm{erg}\ \rm{cm}^{-2}\ \rm{s}^{-1}$), $k$ is the Boltzmann constant, $T$ is the electron temperature, and $n_{\rm{H}}$ is the hydrogen density in ${\rm{cm}}^{-3}$. Taking $F = L / 4\pi r^{2}$ and using $\xi \equiv {L}/{{n_{\rm{H}}\,r^2 }}$, $\Xi$ can be expressed as

\begin{equation}
\label{big_xi_eq_v2}
\Xi  = \frac{L}{{4\pi r^2\, n_{\rm{H}}\, c\, kT}} = \frac{\xi }{{4\pi \, c\, kT}} \approx 19222\ \frac{\xi }{T}
.\end{equation}

On the S curve itself, the heating rate is equal to the cooling rate, so the gas is in thermal balance. To the left of the curve, cooling dominates over heating, while to the right of the curve, heating dominates over cooling. On the branches of the S curve that have a positive gradient, the photoionised gas is thermally stable. This means small perturbations upwards in temperature increase the cooling, while small perturbations downwards in temperature increase the heating. However, on branches with negative gradient, the photoionised gas is thermally unstable. In this case, a small perturbation upwards in temperature increases the heating relative to the cooling, causing further temperature rise, whereas a small perturbation downwards in temperature leads to further cooling. 

\begin{figure}[!tbp]
\centering
\resizebox{1.0\hsize}{!}{\hspace{0cm}\includegraphics[angle=0]{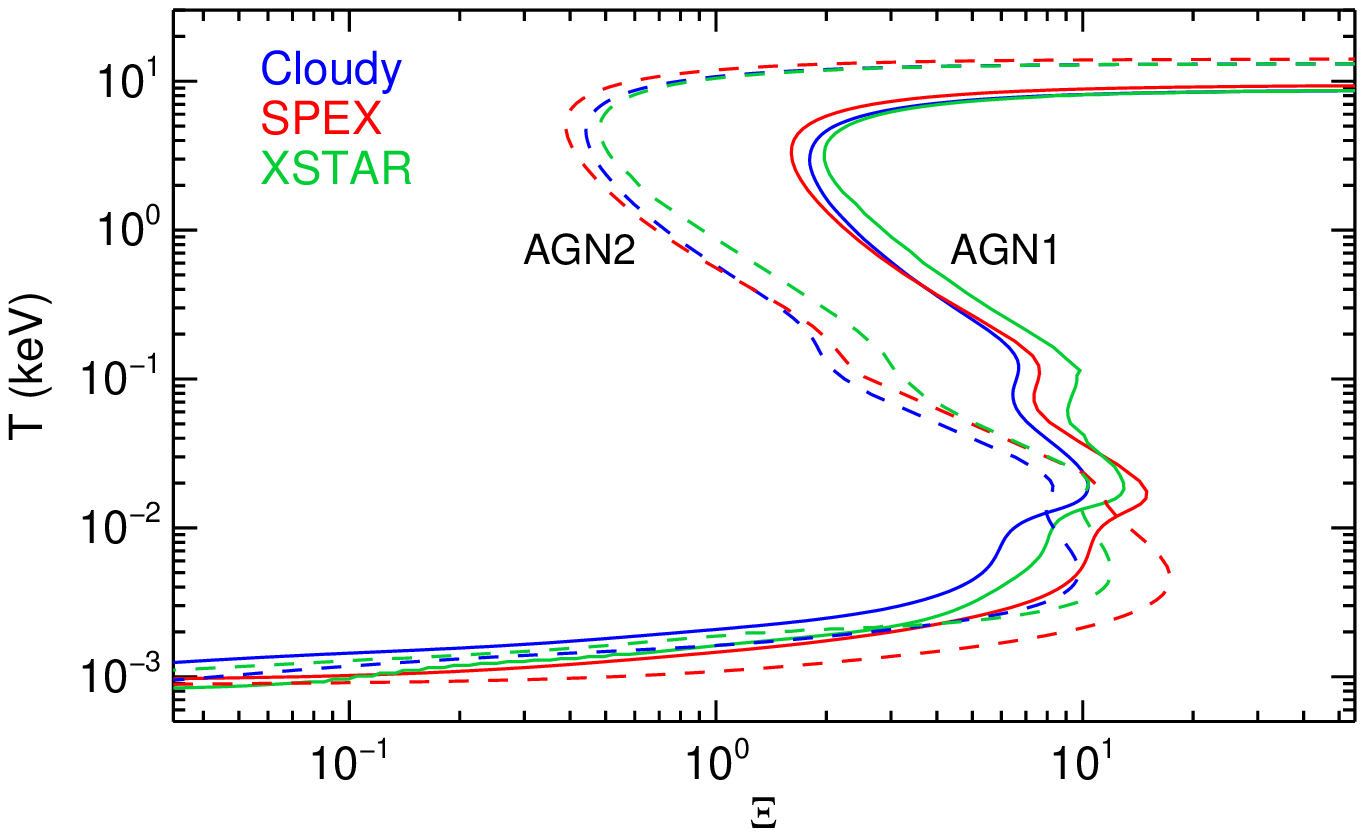}}\vspace{-0.7cm}
\resizebox{1.0\hsize}{!}{\hspace{0cm}\includegraphics[angle=0]{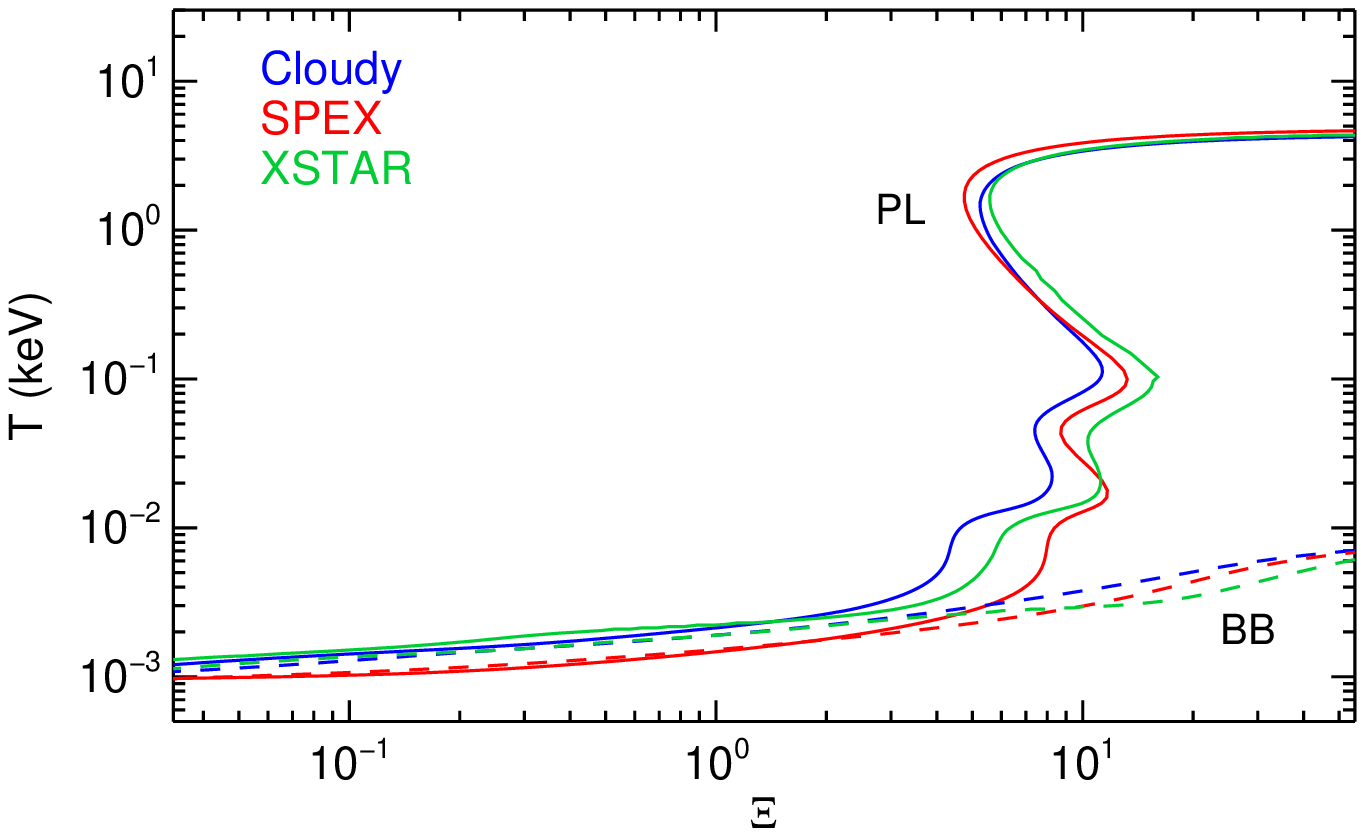}}
\caption{Electron temperature $T$ of a PIE plasma as a function of the pressure form of the ionisation parameter $\Xi$. The curves are calculated using the \cloudy, \spex, and \xstar photoionisation codes, shown in blue, red, and green, respectively. The calculations are carried out for the four different SEDs of Fig. \ref{SED_fig}: AGN1 (top panel in solid line), AGN2 (top panel in dashed line), PL (bottom panel in solid line), and BB (bottom panel in dashed line).}
\label{T_big_Xi}
\end{figure}

In Fig. \ref{T_big_Xi} we show the computed cooling curves, corresponding to the four SED cases of Fig. \ref{SED_fig}. We note that the displayed $\Xi$ range in Fig. \ref{T_big_Xi} corresponds to the same $\xi$ and $T$ range used in Fig. \ref{T_xi}. Comparing the results for AGN1 and AGN2 SEDs, it is clear that the EUV/soft X-ray obscuration has a significant impact on the ionisation balance and thermal stability of the plasma, and produces a more extended unstable branch. For each SED case, the predicted unstable branches from \cloudy, \spex, and \xstar are similar.

%
\begin{table}[!tbp]
\begin{minipage}[t]{\hsize}
\setlength{\extrarowheight}{3pt}
\caption{Comparison of the Compton temperature (\TC) values obtained by the three codes for the four SED cases shown in Fig. \ref{SED_fig}.}
\label{T_C_table}
\centering
\small
\renewcommand{\footnoterule}{}
\begin{tabular}{l| c c c}
\hline \hline
 & \cloudy & \spex  & \xstar  \\
SED & \TC (keV) & \TC (keV) & \TC (keV) \\
\hline

AGN1 & 8.7 & 9.4 & 8.7 \\
AGN2 & 13.1 & 14.1 & 13.1 \\
PL  & 4.6 & 4.8 & 4.5 \\
BB & 0.043 & 0.049 & 0.048 \\

\hline

\end{tabular}
\end{minipage}

\end{table}

\section{Physical processes in photoionised plasmas}
\label{heat_cool_sect}

\begin{figure}[!tbp]
\centering
\resizebox{1.0\hsize}{!}{\hspace{0cm}\includegraphics[angle=0]{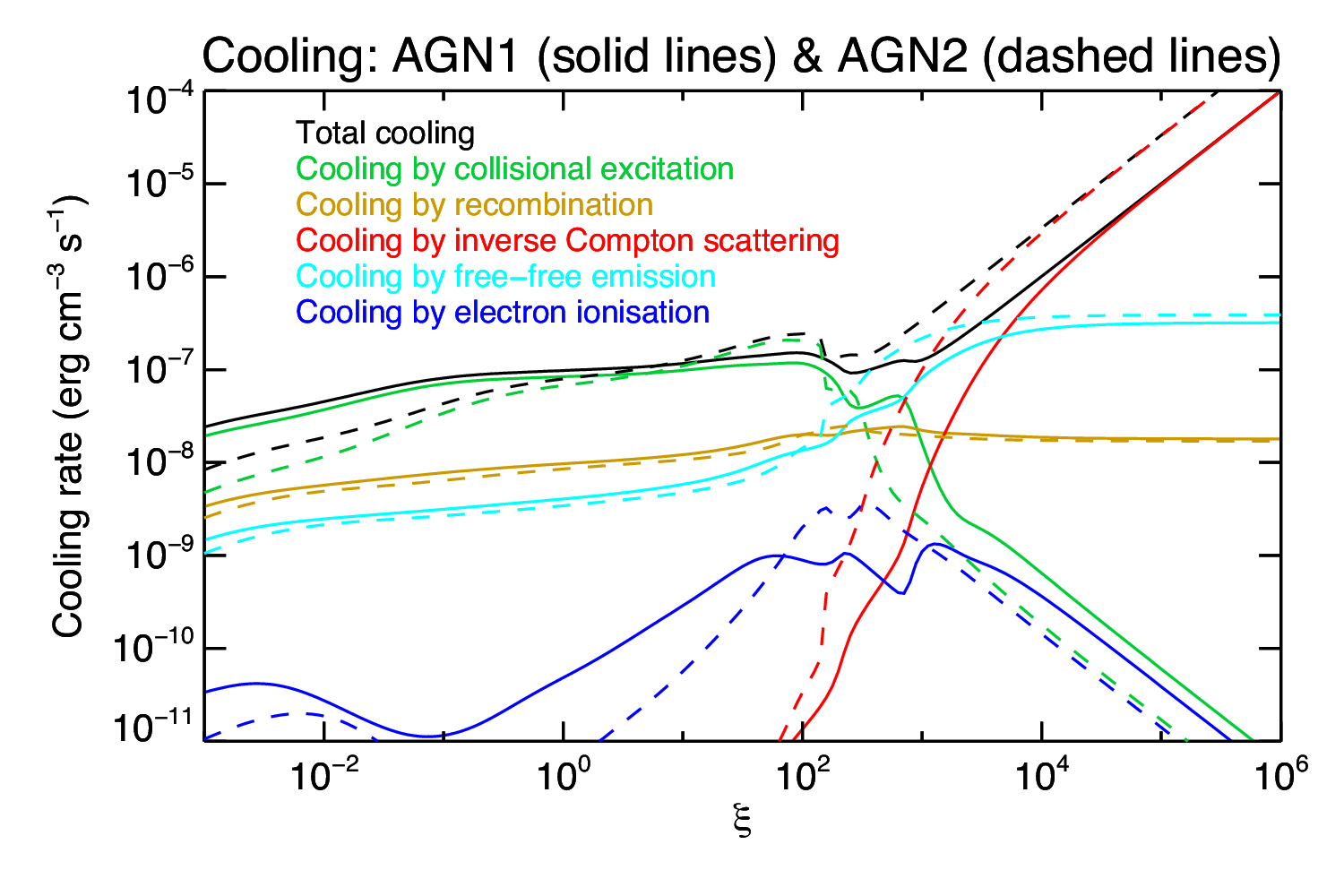}}\vspace{-0.3cm}
\resizebox{1.0\hsize}{!}{\hspace{0cm}\includegraphics[angle=0]{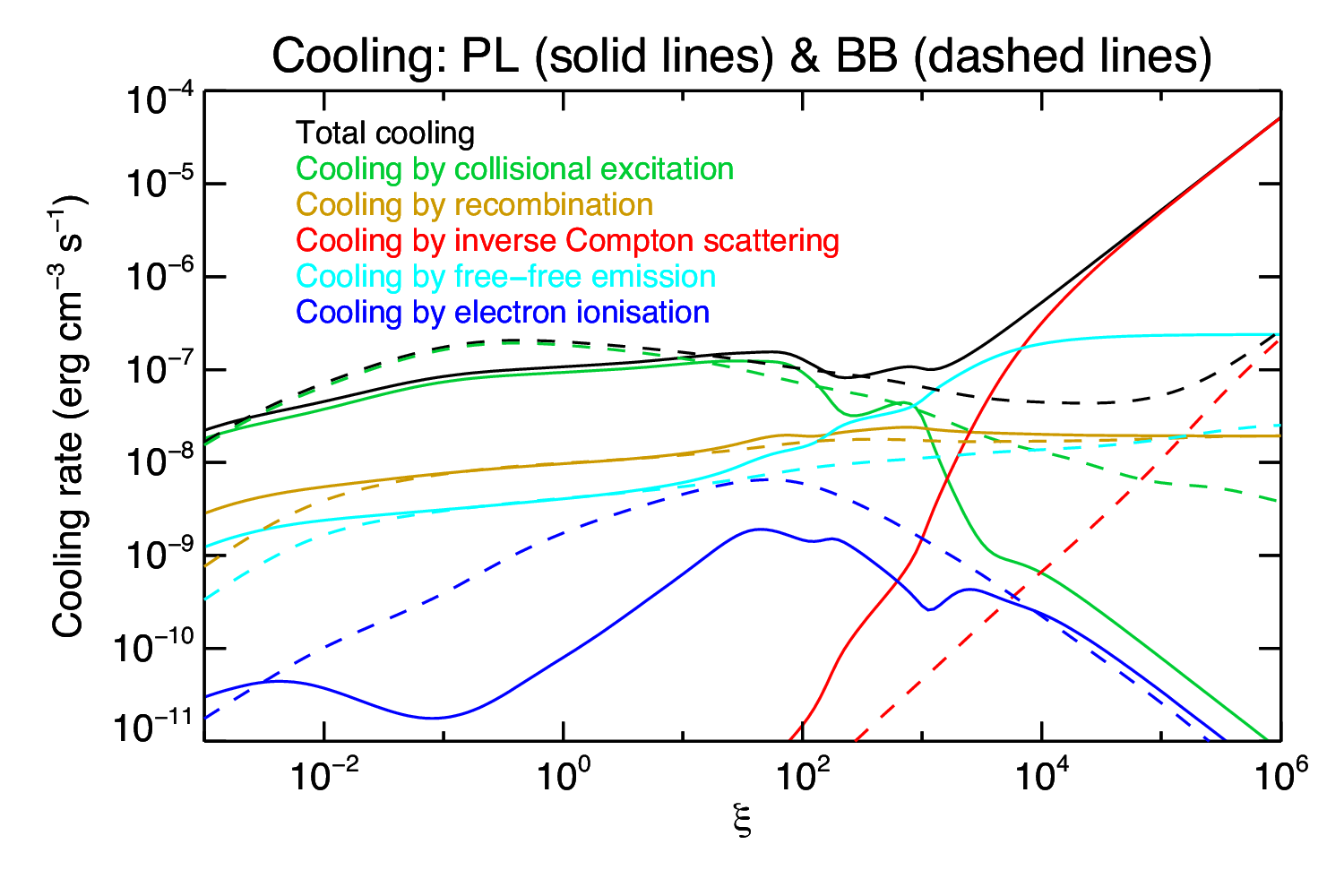}}
\caption{Cooling rate in a PIE plasma as a function of ionisation parameter $\xi$ for the AGN1 and AGN2 SEDs ({\it top panels}), and the PL and BB SEDs ({\it bottom panels}). The curves corresponding to AGN1 and PL are shown in solid lines, and those corresponding to AGN2 and BB are indicated with dashed lines. For each case, the total cooling rate is shown in black, and the contributions from individual processes are shown in the same colours as their corresponding labels.}
\label{cool_rate_fig}
\end{figure}

Figures \ref{heat_rate_fig} and \ref{cool_rate_fig} show how different heating and cooling processes contribute to the total heating and cooling in a PIE plasma. They are derived from our computations using the \spex \pion model for each SED case. They allow us to understand how each process acts under different ionising SEDs, which leads to a different ionisation balance solution. The percentages reported below in our examination of the results, correspond to the fractional contribution by each process to the total heating or cooling rate over the specified $\xi$ range.

%

%

\subsection{Heating processes}

\begin{figure}[!tbp]
\centering
\resizebox{1.0\hsize}{!}{\hspace{0cm}\includegraphics[angle=0]{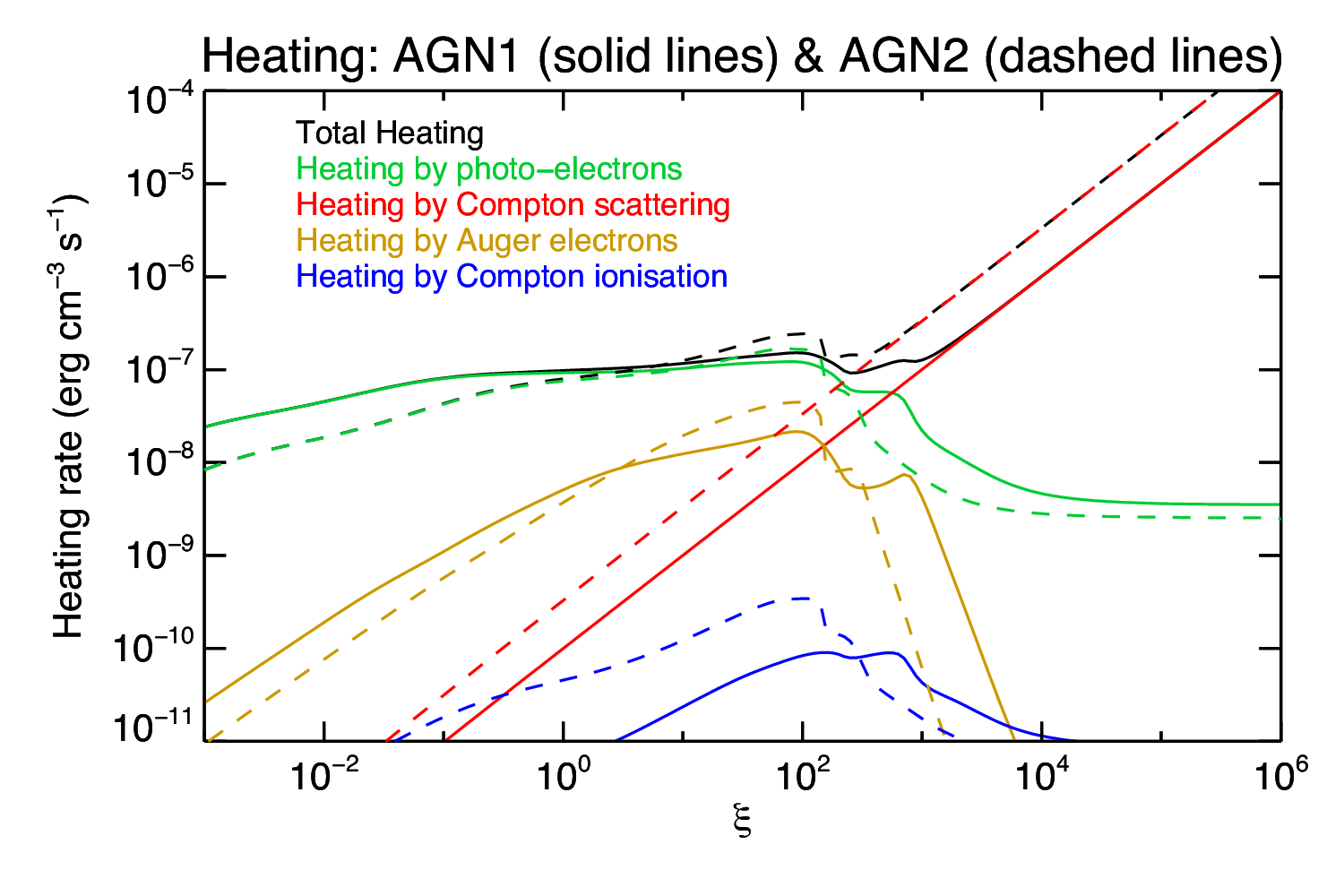}}\vspace{-0.3cm}
\resizebox{1.0\hsize}{!}{\hspace{0cm}\includegraphics[angle=0]{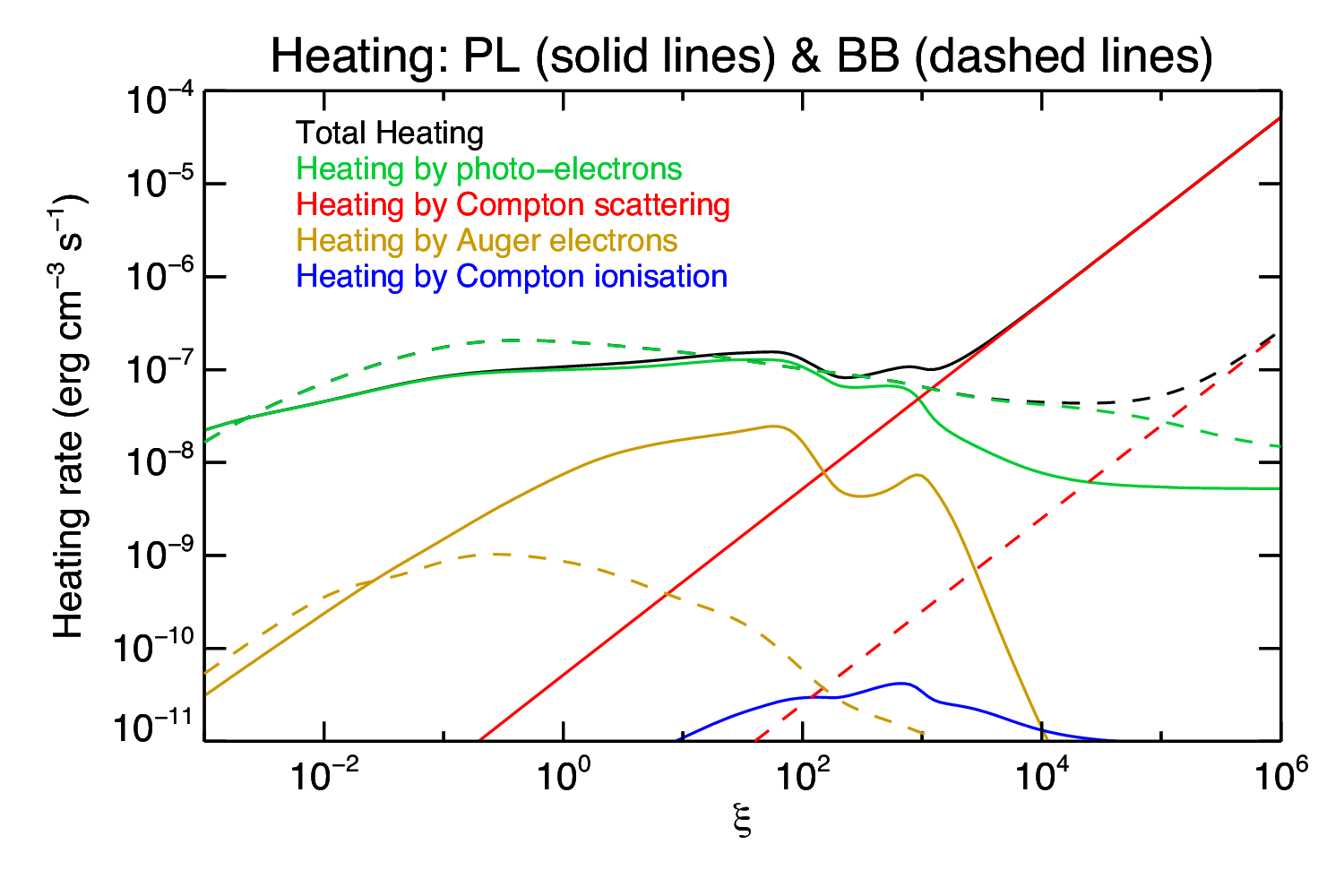}}
\caption{Heating rate in a PIE plasma as a function of ionisation parameter $\xi$ for the AGN1 and AGN2 SEDs ({\it top panels}), and the PL and BB SEDs ({\it bottom panels}). The curves corresponding to AGN1 and PL are shown in solid lines, and those corresponding to AGN2 and BB are indicated with dashed lines. For each case, the total heating rate is shown in black, and the contributions from individual processes are shown in the same colours as their corresponding labels.}
\label{heat_rate_fig}
\end{figure}

From the results of Fig. \ref{heat_rate_fig} we calculated the average heating rate for each process between ${1.0\le \log \xi \le 3.0}$, which is the most relevant ionisation range. We find for all four SED cases, the heating by photo-electrons is the most dominant heating process. For both AGN1 and AGN2 SEDs, the heating processes from strongest to weakest are: (1) photo-electrons, (2) Compton scattering, (3) Auger electrons, and (4) Compton ionisation. For the AGN1 SED, the fractional contribution of these processes to the total heating rate is 72.1\%, 17.7\%, 10.2\%, and 0.05\%, respectively.  For the AGN2 SED, although the order is the same, the values are different. In this case, heating by photo-electrons is lower at 49.7\%, while heating by Compton scattering is higher at 38.9\%. For AGN2, heating by Auger electrons and Compton ionisation are only slightly higher than in AGN1 with values of 11.3\% and 0.09\%, respectively. These differences arise from the significant suppression of EUV/soft X-ray part of the SED in AGN2 relative to AGN1 (see Fig. \ref{SED_fig}).

For the PL SED, the strength of the processes are similar to those of AGN1: 79.1\% for photo-electrons, 9.5\% for Compton scattering, 11.4\% for Auger electrons, and 0.02\% for Compton ionisation. For the BB SED, the heating rates of the processes are rather different. Heating by photo-electrons strongly dominates at 99.8\%, while strengths of the other processes are very small at 0.1\% for Auger electrons, 0.05\% for Compton scattering, and almost zero for Compton ionisation.

Apart from the aforementioned heating processes, the \spex \pion model also takes heating by free-free absorption into account. However, the contribution of this process to the total heating rate is minute and below the displayed range of Fig. \ref{heat_rate_fig}. For ${1.0\le \log \xi \le 3.0}$, the average heating rate by free-free absorption is ${9.5\times10^{-9}}$\% at its lowest point for the BB SED and ${6.5\times10^{-4}}$\% at its highest point for the PL SED.

%

%

\subsection{Cooling processes}

From the results of Fig. \ref{cool_rate_fig} we obtained the average cooling rate for each process between ${1.0\le \log \xi \le 3.0}$. We find that for all four SED cases, cooling by collisional excitation is the most dominant cooling process. For both AGN1 and PL SEDs, the cooling processes ordered from strongest to weakest are as follows: (1) collisional excitation, (2) free-free emission, (3) recombination, (4) electron ionisation, and (5) inverse Comptonisation. For the AGN1 SED, the fractional contribution of these processes to the total cooling rate is 66.8\%, 17.4\%, 14.9\%, 0.6\% and 0.3\%, respectively. For the PL SED, they are 67.3\%, 15.9\%, 15.7\%, 1.0\% and 0.1\%. However, the order and strength of the processes are different for the AGN2 SED. In this case, the contributions from collisional excitation and recombination are lower at 57.6\% and 9.2\%, respectively. On the other hand, the contributions from free-free emission, inverse Comptonisation, and electron ionisation are higher at 27.4\%, 5.0\%, and 0.8\%, respectively. 

For the BB SED, cooling by collisional excitation is higher than those of the other SEDs at 72.7\% of the total cooling rate. Unlike the other SEDs, cooling by recombination is the second strongest process for the BB SED at 14.9\%. Furthermore, cooling by electron ionisation is higher than those of the other SEDs at 4.4\%. On the other hand, cooling by free-free emission and inverse Comptonisation are lower than those of the other SEDs at 7.9\% and 0.01\%, respectively. 

\section{Ionisation state of photoionised plasmas}
\label{ion_frac_sect}

Here we present the ionisation state of PIE plasmas from computations by \cloudy, \spex, and \xstar. We use the AGN1 SED (see Fig. \ref{SED_fig}) for these calculations, which represents the most realistic ionising SED for a typical AGN. We derive the variation of ionic abundances with $\xi$, and compare the temperature $T_{\rm peak}$ and ionisation parameter $\xi_{\rm peak}$ at which ionic abundances peak. 

In Fig. \ref{ionic_fract} we show the ionic fractions of the most relevant ions as a function of $\xi$ in a PIE plasma, calculated by \cloudy, \spex, and \xstar.  To examine the results in Fig. \ref{ionic_fract}, we introduce ${\Delta \log \xi_{\rm peak}}$ and ${\Delta f_{\rm peak}}$, which are defined as the difference between the lowest and highest values of $\log \xi_{\rm peak}$ and $f_{\rm peak}$, respectively, as found by the codes for each ion. For example, we can see in Fig. \ref{ionic_fract} that there is a good agreement between the codes for \ion{O}{vii} and \ion{O}{viii} ions. For each of these ions, $\xi_{\rm peak}$ and $f_{\rm peak}$ values calculated from the codes are close to each other: ${\Delta \log \xi_{\rm peak} \lesssim 0.10}$ and ${\Delta f_{\rm peak} \lesssim 0.05}$. However, towards lower ionisation stages of O, the difference tends to become larger. A similar trend is also found for Fe, where there is better agreement between the codes for high-ionisation ions than low-ionisation ions. We find that from \ion{Fe}{xviii} to \ion{Fe}{xxvi}, ${{\Delta \log \xi_{\rm peak}} \lesssim 0.1}$. From \ion{Fe}{ix} to \ion{Fe}{xvii}, ${{\Delta \log \xi_{\rm peak}} \lesssim 0.2}$. However, towards lower ionised ions, the difference becomes increasingly greater, ranging from 0.3 at \ion{Fe}{viii} to 4 at \ion{Fe}{ii}. The ${\Delta f_{\rm peak}}$ is ${\lesssim 0.05}$ for all ions between \ion{Fe}{ix} and \ion{Fe}{xxvi}, with the exception of \ion{Fe}{xvii}, which is higher at 0.2. For ions between \ion{Fe}{ii} and \ion{Fe}{viii}, ${\Delta f_{\rm peak}}$ ranges between 0.05 and 0.5.

The comparison for partially ionised ions of the most abundant elements are provided in Table \ref{ionic_table} of Appendix \ref{appendix}. In this table we list the temperature $T_{\rm peak}$ and ionisation parameter $\xi_{\rm peak}$ for each partially ionised ion. The ionic fraction value at its peak for each ion is given by $f_{\rm peak}$ in the table. We computed these values for the AGN1 SED using \cloudy, \spex and \xstar. Taking into account all the 177 ions reported in Table \ref{ionic_table}, the mean and median ${\Delta \log \xi_{\rm peak}}$ are 0.44 and 0.16, respectively. The mean and median ${\Delta f_{\rm peak}}$ are 0.11 and 0.05, respectively. The median values provide a better representation because of the few outliers in the list. In Sect. \ref{discussion}, we discuss the deviations between the results of the three codes, given in Table \ref{ionic_table}.

%
\begin{figure*}[!tbp]
\centering
\resizebox{0.87\hsize}{!}{\hspace{0cm}\includegraphics[angle=0]{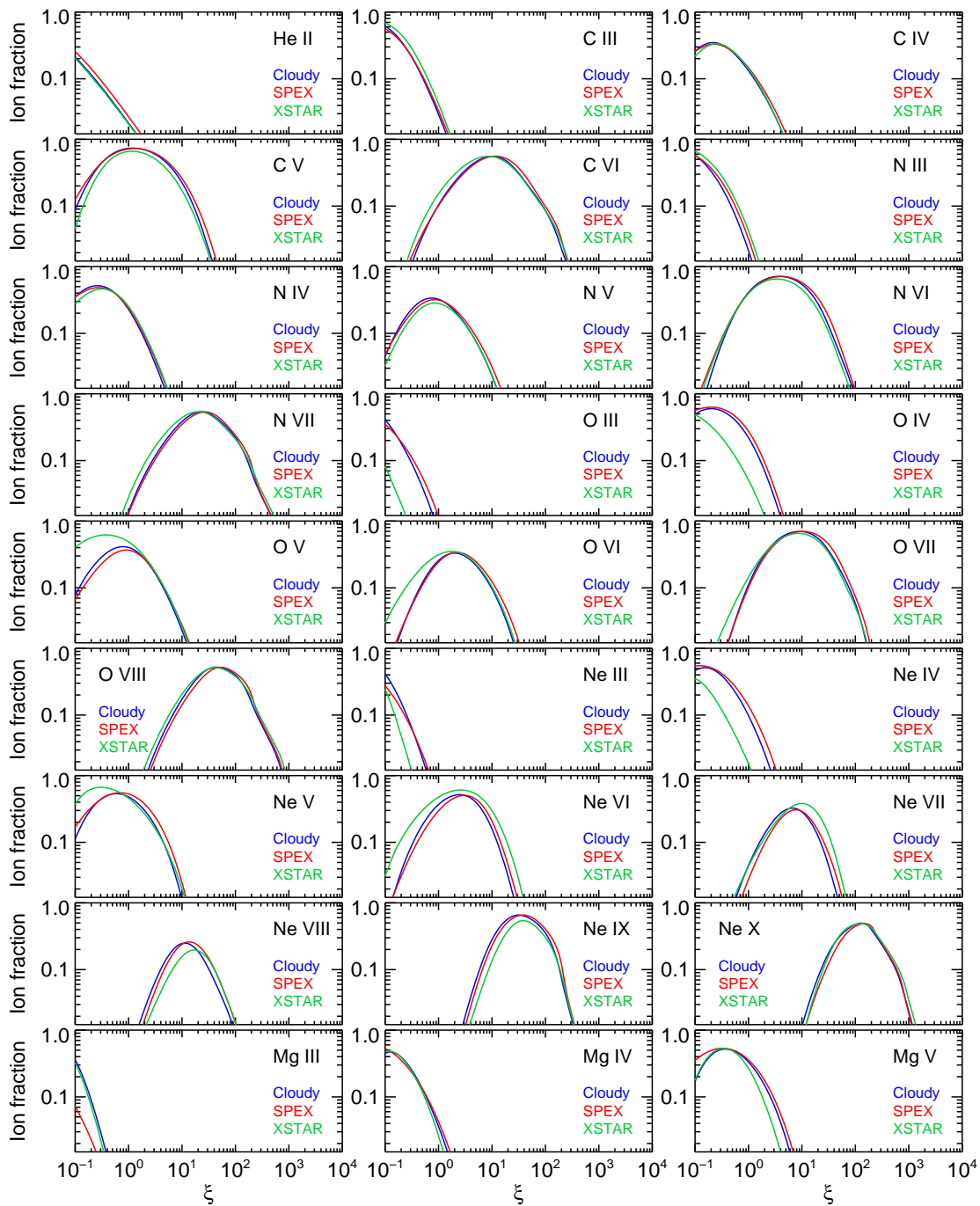}}\vspace{-1.9cm}
\caption{Ionic abundances in a PIE plasma as a function of ionisation parameter $\xi$, which is calculated using \cloudy (shown in blue), \spex (shown in red), and \xstar (shown in green) for the AGN1 SED as described in Sect. \ref{ion_frac_sect}. \textit{Figure continued next page.}}
\label{ionic_fract}
\end{figure*}
\begin{figure*}[!tbp]
\setcounter{figure}{5}
\centering
\resizebox{0.87\hsize}{!}{\hspace{0cm}\includegraphics[angle=0]{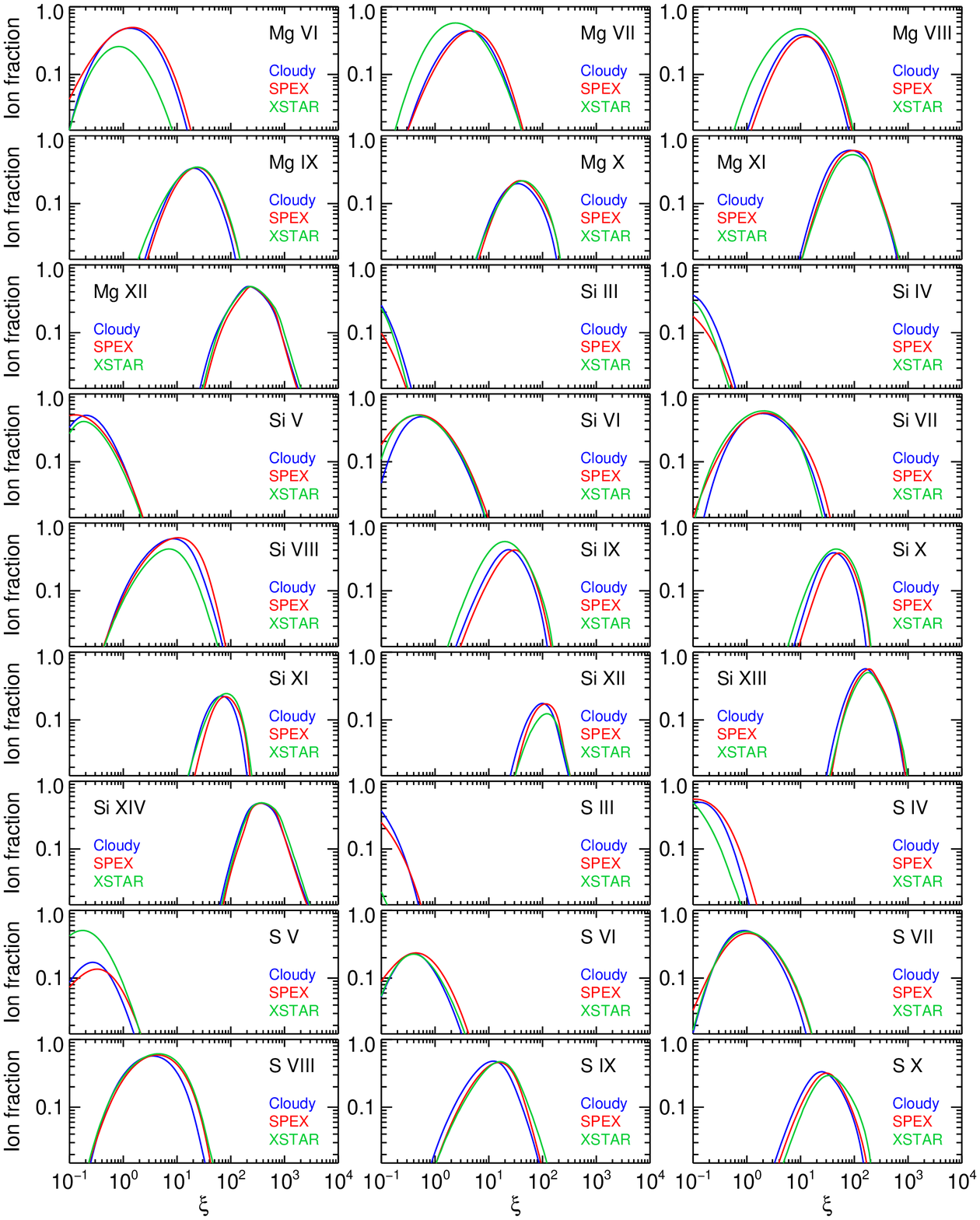}}\vspace{-1.9cm}
\caption{\textit{Figure continued from previous page and continued next page.}}
\end{figure*}
\begin{figure*}[!tbp]
\setcounter{figure}{5}
\centering
\resizebox{0.87\hsize}{!}{\hspace{0cm}\includegraphics[angle=0]{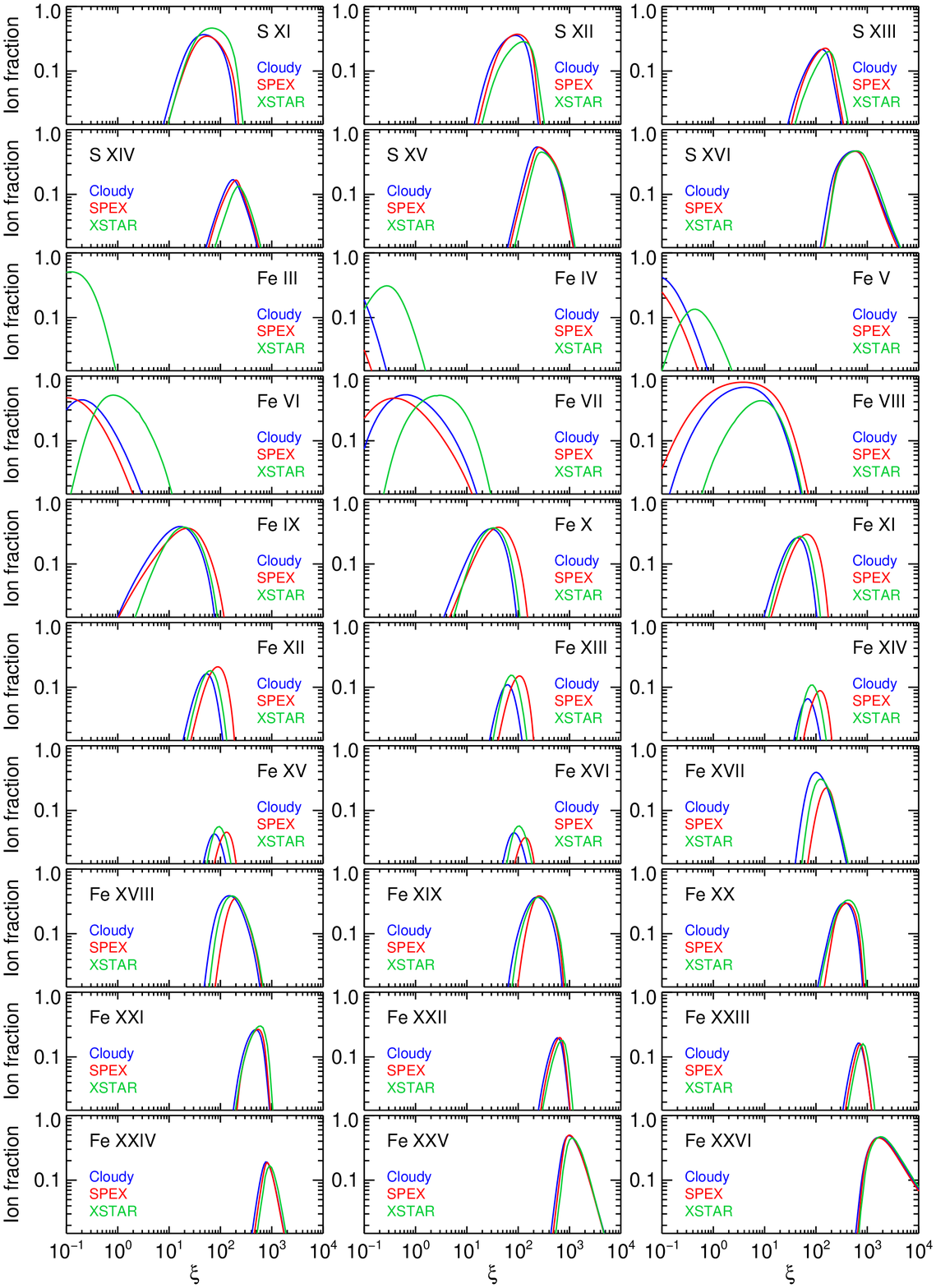}}
\caption{\textit{Figure continued from previous page.}}
\end{figure*}

\section{Transmission of photoionised plasmas in X-rays}
\label{spectra_sect}

\begin{figure}[!tbp]
\centering
\resizebox{1.0\hsize}{!}{\hspace{0cm}\includegraphics[angle=0]{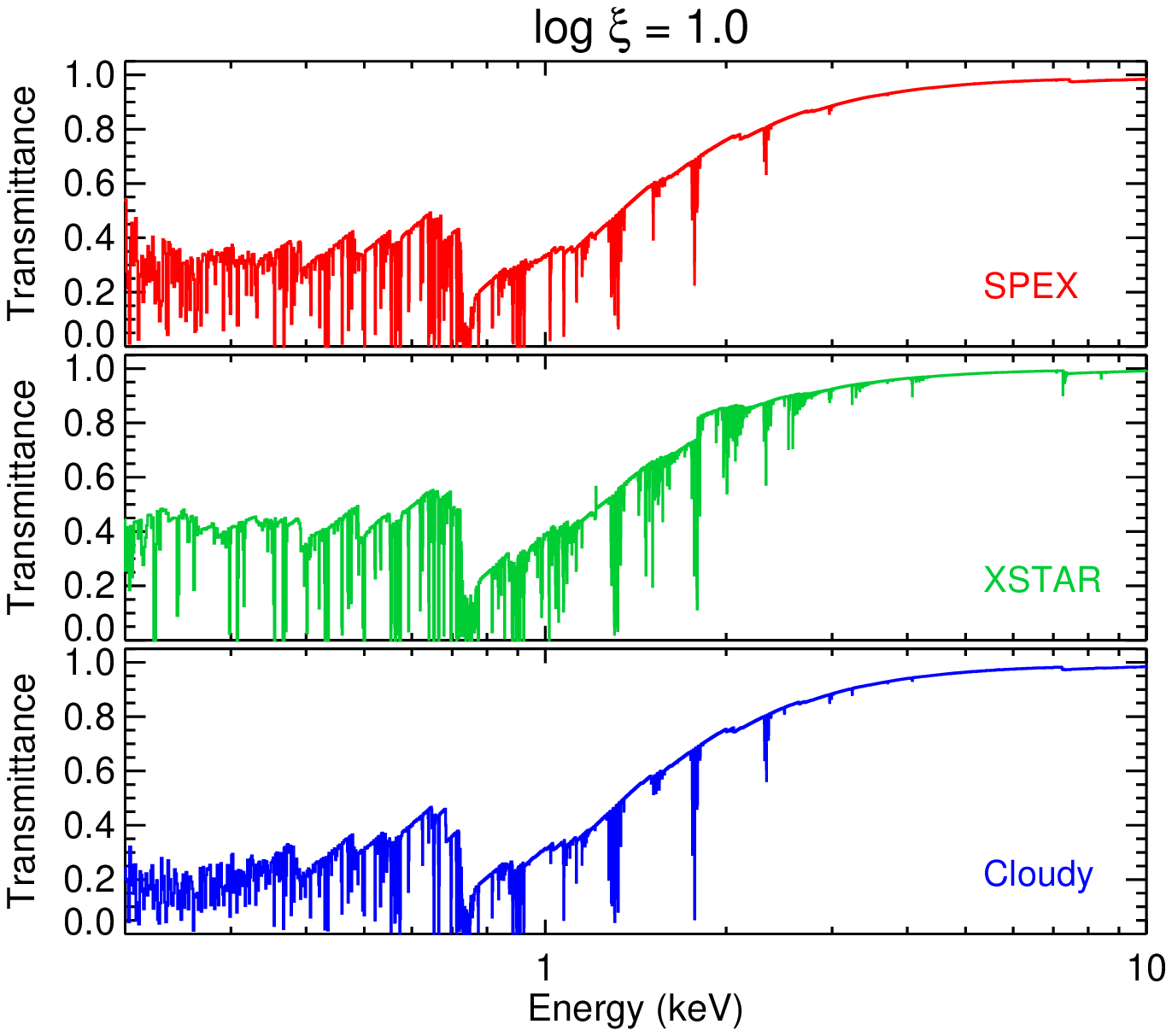}}
\resizebox{1.0\hsize}{!}{\hspace{0cm}\includegraphics[angle=0]{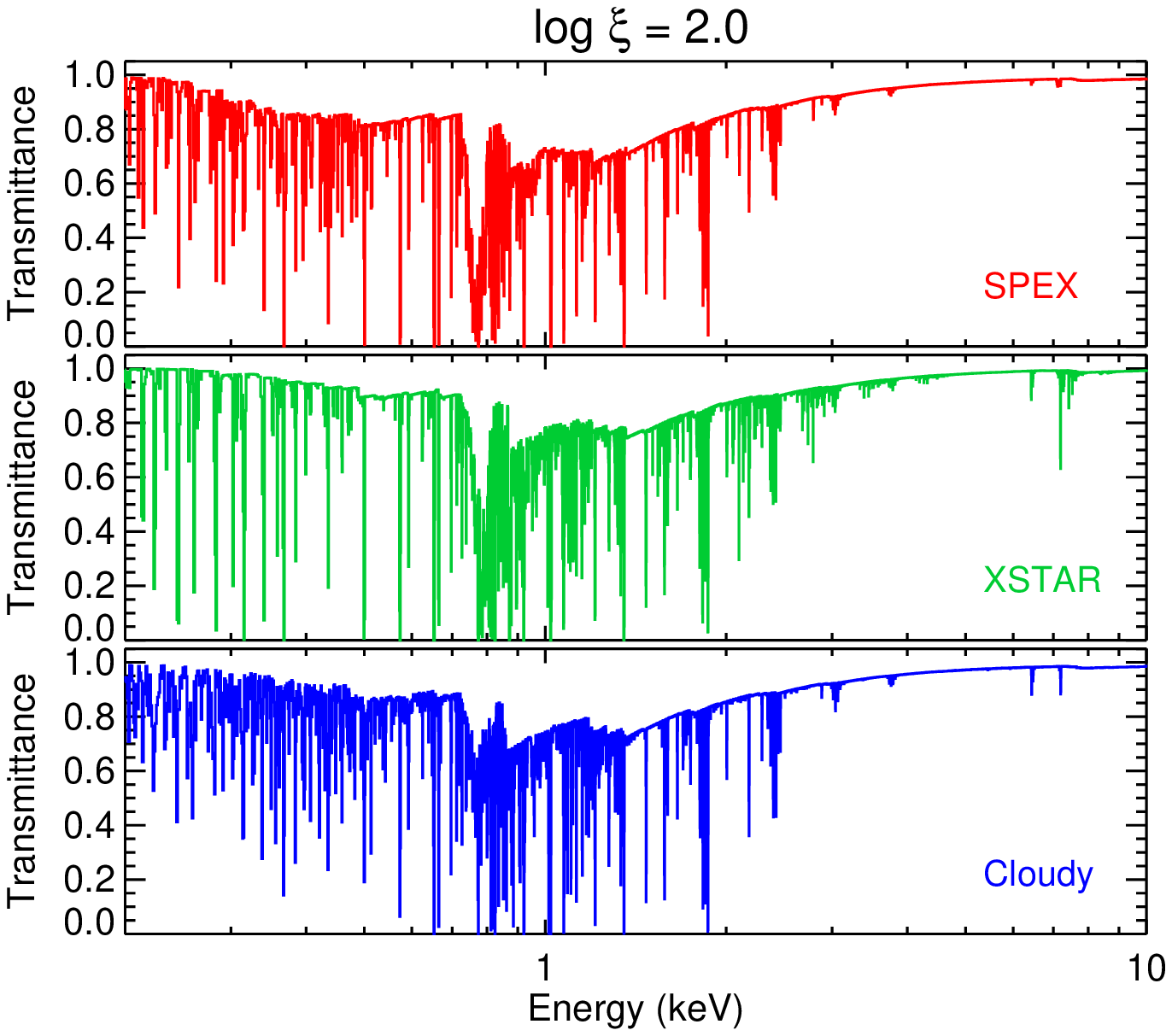}}
\resizebox{1.0\hsize}{!}{\hspace{0cm}\includegraphics[angle=0]{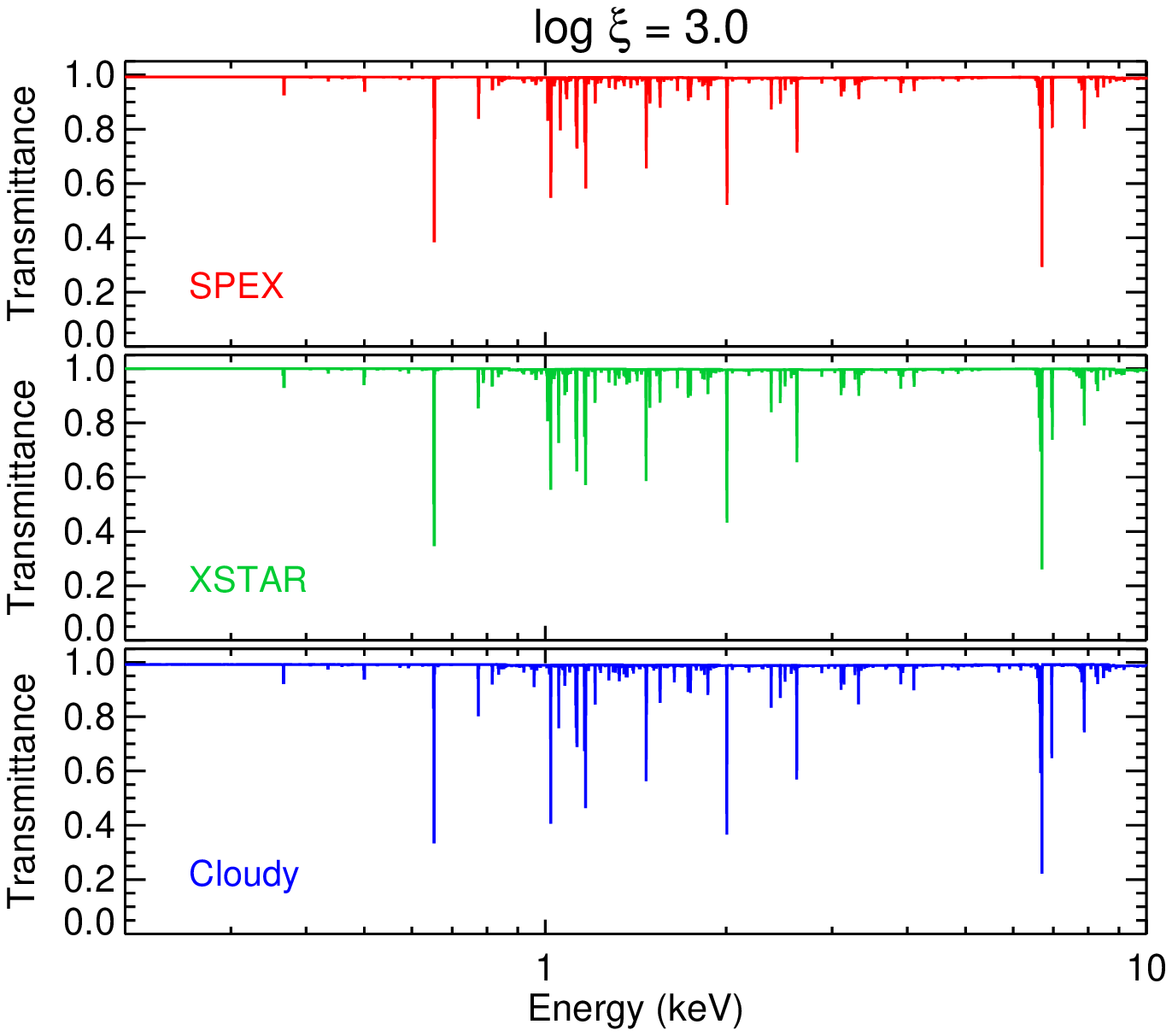}}
\caption{Model transmission X-ray spectra of a PIE plasma with ${\NH = 1 \times 10^{22}\ \rm{cm}^{-2}}$ and $\sigma_{\rm v} = 200\ \kms$, produced by \cloudy, \spex, and \xstar at $\log \xi$ of 1.0, 2.0, and 3.0. The model spectra are calculated for the AGN1 SED as described in Sect. \ref{spectra_sect}.}
\label{spectra_fig}
\end{figure}

Following the computation of ionic abundances for PIE plasmas, we calculated the corresponding X-ray absorption spectra with each code. The absorption spectra were calculated for the AGN1 SED. In our calculations, the column density for a slab of PIE plasma was set to ${\NH = 1 \times 10^{22}\ \rm{cm}^{-2}}$, with a turbulent velocity of ${\sigma_{\rm v} = 200\ \kms}$. These are typical values observed in the AGN ionised outflows. Here we present the results over the most relevant range of ionisation parameters, in which prominent lines are produced in the X-ray band. Figure \ref{spectra_fig} shows the model X-ray spectra calculated by \cloudy, \spex, and \xstar for $\log \xi$ of 1.0, 2.0, and 3.0. The spectra are shown in the rest frame with zero outflow velocity.

\begin{figure*}
\centering
\resizebox{1.0\hsize}{!}{\hspace{0cm}\includegraphics[angle=0]{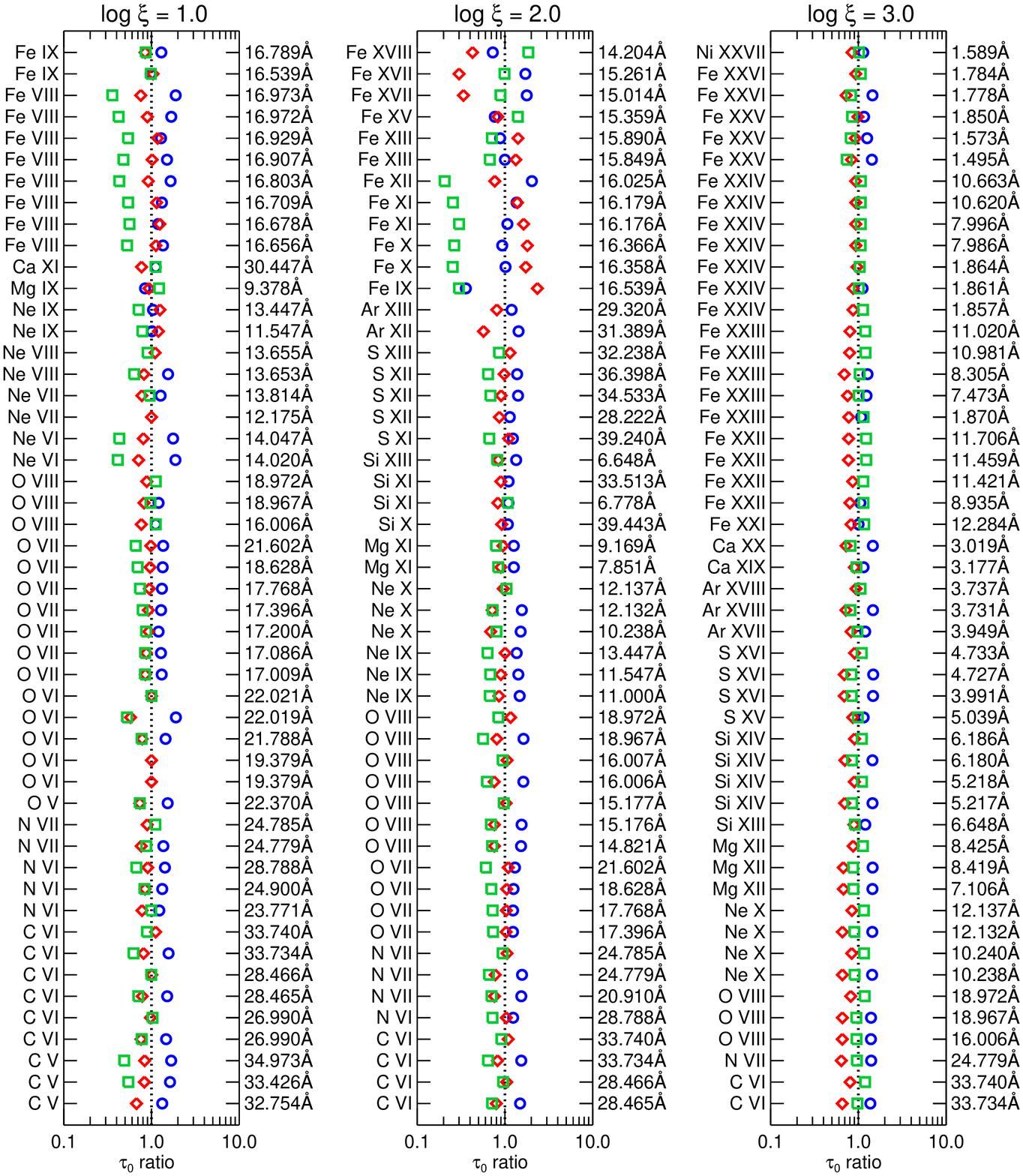}}
\caption{Comparison of the optical depth $\tau_0$ of the strongest X-ray absorption lines at wavelengths below 40~\AA\ (energies above 0.3~keV), from \cloudy, \spex, and \xstar calculations for a PIE plasma, ionised by the AGN1 SED, with ${\NH = 1 \times 10^{22}\ \rm{cm}^{-2}}$ and ${\sigma_{\rm v} = 200\ \kms}$. The displayed data give the ratio of $\tau_0$ from each code relative to the mean $\tau_0$ found by the codes. The $\tau_0$ ratio is shown in blue circles for \cloudy, red diamonds for \spex, and green squares for \xstar. The listed lines correspond to the 50 lines with the highest $\tau_0$ values at $\log \xi$ of 1.0 ({\it left panel}), 2.0 ({\it middle panel}), and 3.0 ({\it right panel}). The dotted vertical line at $\tau_0$ ratio of 1 indicates where the results from the codes would be identical.}
\label{tau0_fig}
\end{figure*}

To compare the end results of our photoionisation calculations, we obtained the optical depth of the strongest X-ray absorption lines. This is useful because the line optical-depth determines the strength of the absorption line in the spectrum. The optical depth at the line centre is given by
\begin{equation}
\label{tau0_eq}
\tau_{0} = \alpha\, h\, \lambda_{\rm c}\, f_{\rm osc}\, N_{\rm ion}\, /\, 2\sqrt{2\pi}\, m_{\rm e}\, \sigma_{{\rm v}}
\end{equation}
where $\alpha$ is the fine structure constant, $h$ the Planck constant, $\lambda_{\rm c}$ the wavelength at the line centre, $f_{\rm osc}$ the oscillator strength, $N_{\rm ion}$ the column density of the absorbing ion, $m_{\rm{e}}$ the electron mass, and $\sigma_{\rm v}$ the velocity dispersion. Thus, by comparing $\tau_0$ of each line from the codes, we are essentially comparing the product of $f_{\rm osc}$ and $N_{\rm ion}$, provided $\lambda_{\rm c}$ and $\sigma_{\rm v}$ are the same. In Fig. \ref{tau0_fig} we present a comparison of the optical depth $\tau_0$ of the strongest X-ray absorption lines at wavelengths below 40~\AA\ (energies above 0.3~keV) from \cloudy, \spex, and \xstar calculations. They were calculated for a PIE plasma, ionised by the AGN1 SED, with ${\NH = 1 \times 10^{22}\ \rm{cm}^{-2}}$ and $\sigma_{\rm v} = 200\ \kms$. The panels show the results for the following three ionisation parameters: $\log \xi$ of 1.0, 2.0, and 3.0. The few missing data points in Fig. \ref{tau0_fig} occur when lines are not found in all the codes. We discuss the comparison results of Fig. \ref{tau0_fig} in Sect. \ref{discussion}.

%

\section{Study of ionised outflows in AGN with high-resolution X-ray spectrometers using different photoionisation codes}
\label{obs_sect}

Here we investigate the impact of using different photoionisation codes on the derived parameters from X-ray observations of PIE plasmas. We simulated spectra of PIE plasmas in AGN with the current and future high-resolution X-ray spectrometers, and obtained the deviation on the model parameters arising from the use of the three photoionisation codes. The simulations were carried out for a PIE plasma ionised by the AGN1 SED (i.e. the \ngc unobscured SED). The column density and turbulent velocity of the plasma were fixed to ${\NH = 1 \times 10^{22}\ \rm{cm}^{-2}}$ and $\sigma_{\rm v} = 200\ \kms$. We carried out the photoionisation calculations and spectral simulations for a range of ionisation values, in which $\log \xi$ ranged between 1.0 and 3.0 with an increment of 0.5. This is a typical range of $\xi$ values seen in X-ray observations of AGN ionised outflows, such as in \ngc and \object{NGC~3783}. For our spectral simulations with the spectrometers, we also included a foreground Galactic interstellar component with ${\NH = 1 \times 10^{20}\ \rm{cm}^{-2}}$ (a typical low Galactic \NH, such as seen in our line of sight towards \ngc), absorbing the AGN1 SED. We used the {\tt hot} model in \spex for modelling the Galactic X-ray absorption component, as described in \citet{Meh15a} for \ngc.

\begin{figure*}
\centering
\resizebox{0.85\hsize}{!}{\hspace{0cm}\includegraphics[angle=0]{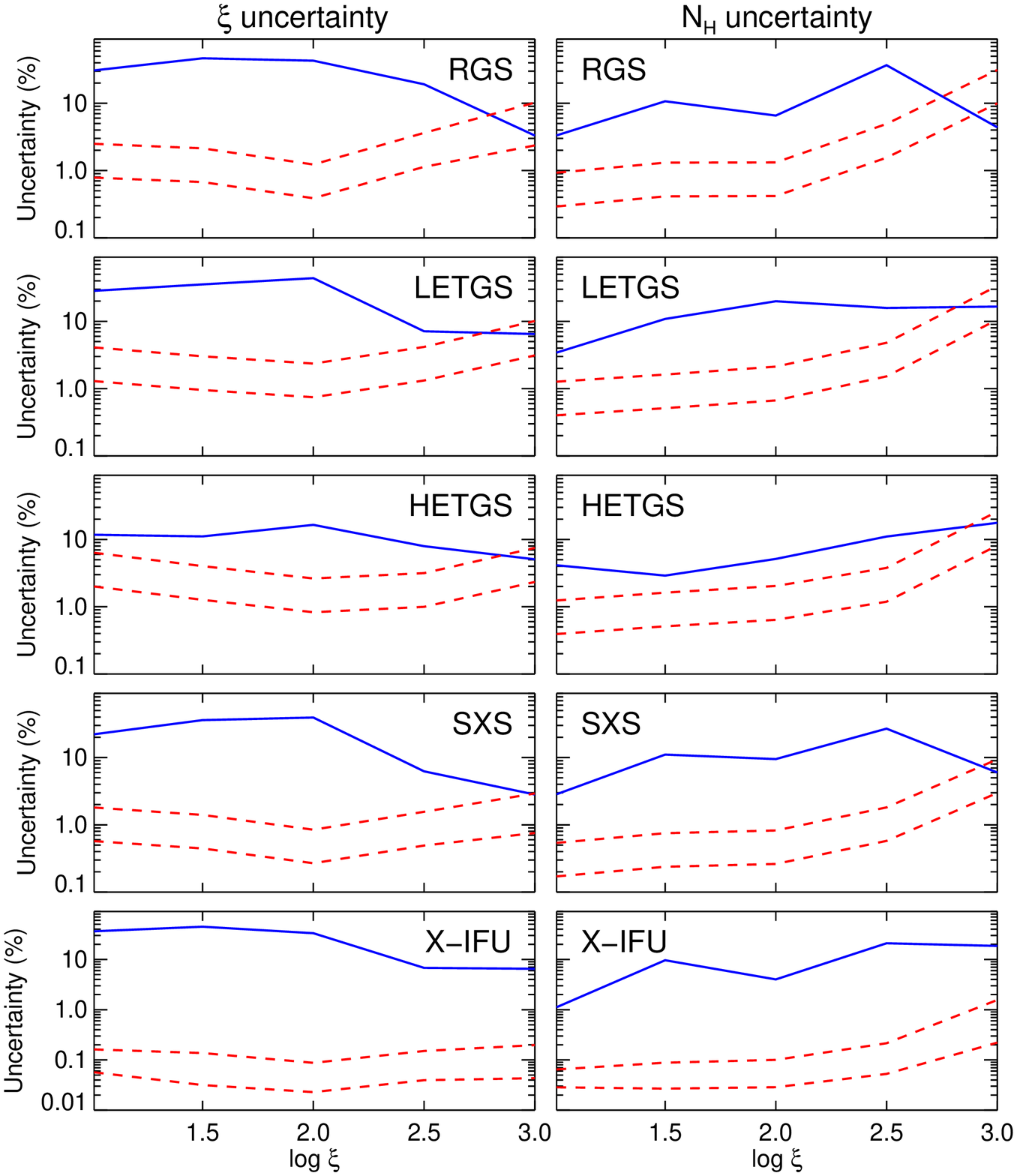}}
\caption{Modelling and observational uncertainties in the derived $\xi$ and \NH parameters of typical ionised outflows in AGN (\ngc). The modelling uncertainties, shown in solid blue lines, correspond to the deviation between the parameter values derived by the \cloudy, \spex, and \xstar photoionisation codes as described in Sect. \ref{obs_sect}. The observational uncertainties, shown in dashed red lines, correspond to the statistical errors of the fitted parameters at 1$\sigma$ confidence level for 100 ks (upper dashed line) and 1 Ms (lower dashed line) simulated observations with each instrument. The left column panels indicate the uncertainties in the $\xi$ parameter, and the right column panels show the uncertainties in the $\NH$ parameter. The photoionisation calculations and spectral simulations were carried out for a PIE plasma ionised by the AGN1 SED (see Fig. \ref{SED_fig}), with $\log \xi$ ranging between 1.0 and 3.0 with an increment of 0.5, ${\NH = 1 \times 10^{22}\ \rm{cm}^{-2}}$ and $\sigma_{\rm v} = 200\ \kms$.}
\label{obs_fig}
\end{figure*}

For each calculation by the three codes, we convolved the corresponding spectrum with the response matrix of each of the high-resolution X-ray spectrometers. We used \xmm RGS \citep{denH01}, \chandra LETGS \citep{Brink00}, and HETGS \citep{Cani05}, \hitomi (\astroh) SXS \citep{Mits14}, and {\it Athena} {X-IFU} \citep{Barr16} for our simulations. For RGS, LETGS, and HETGS, instrumental response matrices from the last observations of \ngc were used, while for SXS and X-IFU we used the latest publicly available response matrices as of September 2016. Each simulated spectrum by a code was fitted with the other two codes to obtain its best-fit $\xi$ and \NH parameters. Thus, any difference in the derived model parameters for a given spectrum would be due to intrinsic differences between the three photoionisation codes. The standard deviation of the fitting results were calculated to represent a measure of the modelling uncertainty in $\xi$ and \NH parameters. For each instrument, we also calculated the observational uncertainties corresponding to the statistical errors of the fitted parameters at 1$\sigma$ confidence level for minimum and maximum exposure times of 100~ks and 1~Ms. The modelling and observational uncertainties are presented in Fig. \ref{obs_fig}. We discuss the deviation in the model parameters obtained by the codes in Sect. \ref{discussion}.

In Fig. \ref{obs_fig}, the modelling uncertainties can be compared with the observational uncertainties. The simulated observational uncertainties for each instrument correspond to the observed X-ray flux level of AGN1 SED (\ngc), which is a bright Seyfert 1 AGN in X-rays with ${F_{{\rm 0.3-2\ keV}}} = 3.1 \times 10^{-11}\ \rm{erg\ cm}^{-2}\ \rm{s}^{-1}$ and ${F_{{\rm 2-10\ keV}}} = 3.5 \times 10^{-11}\ \rm{erg\ cm}^{-2}\ \rm{s}^{-1}$. We can see in Fig. \ref{obs_fig} that the modelling uncertainties are generally larger than the observational uncertainties. However, at ${\log \xi = 3}$, the modelling uncertainties are at the same level or smaller than the observational uncertainties, except for X-IFU, where the observational uncertainties are tiny owing to its exceptional sensitivity. The upcoming \athena X-IFU microcalorimeter will provide us with unprecedented details of the physical structure of the outflowing gas in AGN. In particular, it allows us to accurately determine the properties of the high-ionisation component (${\log \xi = 3}$) of the outflow through the detection of \ion{Fe}{xxiv}, \ion{Fe}{xxv,} and \ion{Fe}{xxvi} lines in the 6 keV band with an energy resolution of 2.5 eV. In the case of AGN1 SED, we find that for a 100 ks X-IFU observation, the statistical errors in the $\xi$ and \NH parameters of the high-ionisation component are smaller than the modelling uncertainties by factors of 30 and 10, respectively.

%

%

\section{Discussion}
\label{discussion}

The deviation in the ionisation state results, derived by the three codes, is presented visually in Fig. \ref{ionic_fract} and numerically in Table \ref{ionic_table} for the most relevant ions. Consequently, any deviation between the codes in ionisation state results in some differences in the strength of the X-ray absorption lines, which are shown in Figs. \ref{spectra_fig} and \ref{tau0_fig}. The effects on the derived plasma parameters, from modelling observational X-ray spectra of AGN ionised outflows, are presented in Fig. \ref{obs_fig}.

In general, the observed differences between the results of the codes are a manifestation of all the little differences in the modelling of the heating/cooling processes and their associated atomic data. Figure \ref{ionic_fract} shows that there is some discrepancy between the codes for the low-ionisation stages of Fe (i.e. \ion{Fe}{i-vii}). This most likely originates from differences in the low-temperature dielectronic recombination (DR) calculations by the codes. In \spex, the ionisation balance calculations of \citet{Brya09} are adopted. The DR and radiative recombination (RR) rate coefficients used in \citet{Brya09} are the same as in \citet{Brya06}, but these data are updated to include corrections to some of the rate coefficients, as well as updated DR data for Mg-like ions of H through Zn, and for Al-like to Ar-like ions of Fe, taken from \citet{Badn06a, Badn06b, Altu07}. The DR and RR data for all other ions, including \ion{Fe}{i-vii}, are from \citet{Mazz98}. The rates from \citet{Mazz98} do not include the low-temperature DR for \ion{Fe}{i-vii}. However, in \cloudy, the low-temperature DR for \ion{Fe}{i-v} is estimated using the mean of all the existing DR rates for each ionisation stage, while for \ion{Fe}{vi-vii}, the DR rates calculated by M.F. Gu are used (private communication). On the other hand, in \xstar, the low-temperature DR for \ion{Fe}{i-viii} is estimated from the low-temperature part of the \ion{Fe}{ix} DR rate. Both \cloudy and \xstar use the DR rates of \citet{Badn06a} for \ion{Fe}{ix} to \ion{Fe}{xiii}, as well as the DR rates provided online\footnote{\url{http://amdpp.phys.strath.ac.uk/tamoc/DR/}} by N.R. Badnell for higher ionisation stages of Fe.

There is about a 4\% deviation in the value of Compton temperatures obtained by the three codes. Some of this deviation is attributed to the relativistic corrections used by the codes for the energy exchange by relativistic electrons in Compton scattering. \cloudy uses numerical fits to the results of \citet{Wins75}, as used in \citet{Kro81}, which were provided by C.B. Tarter (private communication). The treatment of Compton heating and cooling in \xstar versions prior to 2.3 were not accurate for hard spectra with significant flux above 100 keV. However, this has been updated in version 2.3 and this paper, using rates from I. Khabibullin (private communication), based on the expressions given by \citet{Shes88}. The energy shift per scattering is calculated by interpolating in a table. In \spex, the heating/cooling by Compton scattering is calculated using the formulae provided by \citet{Fer88} (originally from \citealt{Wins75}) and \citet{Levi70}.

From analysis of the ionic fractions of partially ionised ions in Table \ref{ionic_table}, we find that for H-like and He-like sequence ions there is on average about a 7\% deviation between the codes in $\xi_{\rm peak}$, at which ionic fractions peak. This deviation between the codes rises slightly to 9\% for Li-like ions, 11\% for Be-like ions, and 13\% for B-like ions. The deviation becomes greater for higher isoelectronic sequence ions: 39\% on average for the C-like through Fe-like ions listed in Table \ref{ionic_table}. If one considers all the ions in the table, the average deviation in $\xi_{\rm peak}$ between the codes is about 28\%.

By comparing the optical depth $\tau_0$ of the X-ray absorption lines shown in Fig. \ref{tau0_fig}, we find that on average there is about 30\% deviation between the codes for lines produced at $\log \xi$ of 1 and 2. This deviation is reduced to 20\% at $\log \xi$ of 3. In general, there is better agreement between the codes for higher ionisation ions than their lower ionisation counterparts. Although the above comparison of ionic fractions and optical depths applies to the AGN1 ionising SED, similar deviations between the results of the codes are also found for the other three SEDs. The optical depth $\tau_0$ of a line depends on the column density of the corresponding ion ($N_{\rm ion}$), oscillator strength $f_{\rm osc}$ of the line transition, and its wavelength at the line centre $\lambda_{\rm c}$ (see Eq. \ref{tau0_eq}). These parameters can be potentially different between the codes for a given line. However, the observed differences in $\tau_0$ between the codes (Fig. \ref{tau0_fig}) is predominantly caused by $N_{\rm ion}$, which is determined by the ionisation balance calculation from each code. We find that small differences between the codes in $f_{\rm osc}$ or $\lambda_{\rm c}$, have negligible contributions to the deviations in the results.

Finally, we considered a practical application of the comparison between the codes in Sect. \ref{obs_sect}, where the impact on the modelling of the observational spectra of AGN ionised outflows were examined. The results of Fig. \ref{obs_fig} show that the modelling uncertainty in $\xi$ stays relatively unchanged between $\log \xi$ of 1.0 and 2.0 at a level of 30\% on average for all the instruments. Towards higher ionisation, the $\xi$ modelling uncertainty becomes smaller. Between $\log \xi$ of 2.5 and 3.0, the modelling uncertainty is 7\% on average. Furthermore, the modelling uncertainty in \NH appears to generally increase from low to high ionisation; this uncertainty is 3\% on average at ${\log \xi = 1.0}$ and 22\% at ${\log \xi = 2.5}$. However, at ${\log \xi =3}$, the modelling uncertainty in \NH reduces to 13\% on average. For spectroscopic study of ionised outflows in AGN, such levels of modelling uncertainty in the ionisation parameter and column density, indicate that the intrinsic differences in the codes would not greatly alter our scientific interpretations of X-ray spectra.

\section{Conclusions}
\label{conclusions}
We have carried out a systematic comparison of results from photoionisation calculations with the \cloudy, \spex, and \xstar codes. From the findings of our investigation we conclude the following:

\begin{enumerate}
\item In general, there is reasonable agreement between the codes for the thermal and ionisation states that they derive, in particular for high-ionisation plasmas. There is about 10\% deviation between the codes in the ionisation parameter $\xi$ at which ionic abundances of H-like to B-like ions peak. For higher isoelectronic sequence ions, the deviation becomes larger at about 40\% on average for C-like to Fe-like ions. The deviation in $\xi$ for all ions is about 30\% on average.
\item The Compton temperature values calculated by the three codes deviate by about 4\% from each other for the various ionising SED cases that we have investigated.
\item The computed optical depth $\tau_0$ of the strongest X-ray absorption lines from photoionised plasma in AGN deviates between the codes by about 30\% for lines produced at ${\log \xi}$ of 1 to 2, and decreases to about 20\% for lines at ${\log \xi}$ of 3. This deviation in $\tau_0$ is predominately caused by differences in the derived ionic column densities, rather than being due to differences in the value of the atomic parameters used in the codes.
\item From spectral simulations of AGN ionised outflows with \xmm RGS, \chandra LETGS and HETGS, \hitomi SXS, and \athena X-IFU, we find that there is about 10--40\% deviation between the different photoionisation codes in their derived values for the model parameters of the outflows. Such levels of modelling differences are unlikely to greatly impact the scientific interpretation of the observed X-ray spectra of AGN outflows.

\item The observational uncertainties on the best-fit parameters of photoionised plasmas in X-ray bright AGN are generally smaller than the corresponding modelling uncertainties arising from different photoionisation codes. Our results highlight the importance of continuous development and enhancement of the models and atomic data, which are incorporated in the photoionisation codes, in particular for the upcoming era of X-ray astronomy with \athena.
\end{enumerate}

\begin{acknowledgements}

SRON is supported financially by NWO, the Netherlands Organization for Scientific Research. We thank Gary Ferland for useful discussions. We thank the anonymous referee for useful comments.

\end{acknowledgements}


\appendix

\section{Ionic abundances in photoionised plasmas}
\label{appendix}

In Table \ref{ionic_table} we list the temperature $T_{\rm peak}$ and ionisation parameter $\xi_{\rm peak}$, at which ionic abundances of partially ionised ions peak. They were computed for the AGN1 ionising SED (Fig. \ref{SED_fig}) using the \cloudy, \spex, and \xstar codes. The ionic fraction value for each ion at its peak is given by $f_{\rm peak}$ in the table.

%
\begin{longtab}
\clearpage
\onecolumn
\small
\setcounter{LTchunksize}{10}
\setlength{\LTcapwidth}{5.4in}
\begin{longtable}{l | l l l | l l l | l l l}
\caption{\label{ionic_table} Temperature $T_{\rm peak}$ and ionisation parameter $\xi_{\rm peak}$, at which abundance of each partially ionised ion in a photoionised plasma peaks. The ionic fraction at the peak is given by $f_{\rm peak}$ for each ion. These are computed for the AGN1 SED (Fig. \ref{SED_fig}) using the \cloudy, \spex, and \xstar codes as described in Sect. \ref{ion_frac_sect}.}\\
\hline\hline
 &  \multicolumn{3}{c|}{$T_{\rm peak}$ (eV)} &  \multicolumn{3}{c|}{$\log \xi_{\rm peak}$}  &  \multicolumn{3}{c}{Peak ionic fraction $f_{\rm peak}$}  \\
Ion & \cloudy & \spex & \xstar & \cloudy & \spex & \xstar & \cloudy & \spex & \xstar \\
\hline
\endfirsthead
\caption{continued.}\\
\hline\hline
 &  \multicolumn{3}{c|}{$T_{\rm peak}$ (eV)} &  \multicolumn{3}{c|}{$\log \xi_{\rm peak}$}  &  \multicolumn{3}{c}{Peak ionic fraction $f_{\rm peak}$}  \\
Ion & \cloudy & \spex & \xstar & \cloudy & \spex & \xstar & \cloudy & \spex & \xstar \\
\hline
\endhead
\hline
\endfoot

He    II     &         0.94 &     0.91 &     0.76 &        -2.43 &    -2.43 &    -2.47 &         0.82 &     0.82 &     0.81 \\
\hline
C     II     &         0.79 &     0.84 &     0.62 &        -3.21 &    -2.99 &    -3.10 &         0.83 &     0.83 &     0.82 \\
C     III    &         1.31 &     1.00 &     0.92 &        -1.44 &    -1.40 &    -1.37 &         0.77 &     0.68 &     0.79 \\
C     IV     &         1.59 &     1.16 &     1.29 &        -0.66 &    -0.65 &    -0.60 &         0.35 &     0.33 &     0.33 \\
C     V      &         2.07 &     1.56 &     1.68 &         0.09 &     0.11 &     0.08 &         0.74 &     0.73 &     0.66 \\
C     VI     &         3.88 &     2.92 &     2.95 &         1.02 &     1.06 &     0.92 &         0.56 &     0.56 &     0.56 \\
\hline
N     II     &         0.90 &     0.86 &     0.69 &        -2.58 &    -2.92 &    -2.81 &         0.93 &     0.79 &     0.91 \\
N     III    &         1.34 &     0.98 &     0.90 &        -1.35 &    -1.59 &    -1.45 &         0.73 &     0.76 &     0.83 \\
N     IV     &         1.63 &     1.20 &     1.33 &        -0.57 &    -0.56 &    -0.48 &         0.52 &     0.48 &     0.46 \\
N     V      &         1.91 &     1.46 &     1.59 &        -0.12 &    -0.07 &    -0.05 &         0.34 &     0.32 &     0.28 \\
N     VI     &         2.62 &     2.05 &     2.10 &         0.57 &     0.61 &     0.54 &         0.71 &     0.71 &     0.65 \\
N     VII    &         6.42 &     4.85 &     4.92 &         1.35 &     1.43 &     1.30 &         0.54 &     0.53 &     0.54 \\
\hline
O     II     &         0.94 &     0.86 &     0.73 &        -2.43 &    -2.91 &    -2.60 &         0.87 &     0.72 &     0.80 \\
O     III    &         1.31 &     0.96 &     0.84 &        -1.44 &    -1.78 &    -1.79 &         0.67 &     0.67 &     0.59 \\
O     IV     &         1.57 &     1.14 &     1.01 &        -0.69 &    -0.70 &    -1.20 &         0.60 &     0.63 &     0.53 \\
O     V      &         1.93 &     1.46 &     1.36 &        -0.09 &    -0.04 &    -0.39 &         0.41 &     0.37 &     0.62 \\
O     VI     &         2.26 &     1.77 &     1.81 &         0.30 &     0.34 &     0.24 &         0.33 &     0.34 &     0.35 \\
O     VII    &         3.63 &     2.78 &     2.95 &         0.96 &     1.01 &     0.92 &         0.70 &     0.70 &     0.65 \\
O     VIII   &        10.84 &     7.85 &     8.98 &         1.65 &     1.72 &     1.64 &         0.52 &     0.52 &     0.51 \\
\hline
Ne    II     &         0.76 &     0.78 &     0.51 &        -3.51 &    -3.61 &    -3.65 &         0.71 &     0.72 &     0.74 \\
Ne    III    &         1.19 &     0.95 &     0.84 &        -1.74 &    -2.01 &    -1.79 &         0.86 &     0.72 &     0.87 \\
Ne    IV     &         1.53 &     1.08 &     1.06 &        -0.81 &    -0.91 &    -1.07 &         0.51 &     0.55 &     0.36 \\
Ne    V      &         1.82 &     1.40 &     1.33 &        -0.24 &    -0.16 &    -0.52 &         0.54 &     0.55 &     0.67 \\
Ne    VI     &         2.36 &     1.93 &     1.95 &         0.39 &     0.49 &     0.41 &         0.52 &     0.50 &     0.61 \\
Ne    VII    &         3.13 &     2.55 &     3.27 &         0.81 &     0.89 &     1.01 &         0.33 &     0.31 &     0.38 \\
Ne    VIII   &         4.03 &     3.23 &     4.36 &         1.05 &     1.13 &     1.22 &         0.25 &     0.26 &     0.20 \\
Ne    IX     &         8.54 &     6.41 &     8.24 &         1.50 &     1.58 &     1.60 &         0.66 &     0.66 &     0.54 \\
Ne    X      &        20.23 &    17.57 &    16.76 &         2.10 &     2.18 &     2.11 &         0.50 &     0.49 &     0.49 \\
\hline
Na    II     &         0.57 &     0.59 &     0.29 &        -5.25 &    -6.78 &    -4.93 &         0.97 &     0.99 &     0.97 \\
Na    III    &         1.14 &     0.93 &     0.83 &        -1.86 &    -2.23 &    -2.00 &         0.84 &     0.65 &     0.70 \\
Na    IV     &         1.47 &     1.03 &     1.19 &        -0.96 &    -1.14 &    -0.86 &         0.50 &     0.61 &     0.63 \\
Na    V      &         1.78 &     1.38 &     1.59 &        -0.30 &    -0.21 &    -0.05 &         0.60 &     0.61 &     0.52 \\
Na    VI     &         2.33 &     1.87 &     2.00 &         0.36 &     0.44 &     0.46 &         0.44 &     0.41 &     0.40 \\
Na    VII    &         3.22 &     2.55 &     2.80 &         0.84 &     0.91 &     0.88 &         0.49 &     0.50 &     0.48 \\
Na    VIII   &         4.74 &     3.63 &     4.36 &         1.17 &     1.25 &     1.22 &         0.30 &     0.30 &     0.32 \\
Na    IX     &         6.42 &     4.85 &     5.60 &         1.35 &     1.46 &     1.39 &         0.20 &     0.21 &     0.20 \\
Na    X      &        11.68 &     9.50 &     8.98 &         1.71 &     1.80 &     1.64 &         0.63 &     0.63 &     0.43 \\
Na    XI     &        34.35 &    26.54 &    15.73 &         2.25 &     2.30 &     2.07 &         0.49 &     0.49 &     0.56 \\
\hline
Mg    II     &         0.49 &     0.80 &     0.64 &        -5.94 &    -3.38 &    -3.06 &         0.97 &     0.30 &     0.40 \\
Mg    III    &         1.19 &     0.93 &     0.85 &        -1.74 &    -2.27 &    -1.71 &         0.89 &     0.67 &     0.86 \\
Mg    IV     &         1.49 &     1.01 &     1.15 &        -0.90 &    -1.32 &    -0.94 &         0.47 &     0.61 &     0.49 \\
Mg    V      &         1.69 &     1.22 &     1.33 &        -0.45 &    -0.48 &    -0.48 &         0.52 &     0.53 &     0.53 \\
Mg    VI     &         2.07 &     1.60 &     1.56 &         0.09 &     0.17 &    -0.09 &         0.48 &     0.49 &     0.26 \\
Mg    VII    &         2.67 &     2.20 &     1.91 &         0.60 &     0.69 &     0.37 &         0.44 &     0.44 &     0.58 \\
Mg    VIII   &         4.03 &     3.06 &     3.27 &         1.05 &     1.09 &     1.01 &         0.39 &     0.36 &     0.47 \\
Mg    IX     &         5.75 &     4.15 &     5.60 &         1.29 &     1.37 &     1.39 &         0.33 &     0.34 &     0.35 \\
Mg    X      &         9.48 &     6.41 &     8.98 &         1.56 &     1.59 &     1.64 &         0.20 &     0.22 &     0.22 \\
Mg    XI     &        14.59 &    12.72 &    14.35 &         1.92 &     1.98 &     1.98 &         0.61 &     0.61 &     0.53 \\
Mg    XII    &        51.65 &    38.05 &    37.14 &         2.34 &     2.37 &     2.36 &         0.48 &     0.48 &     0.48 \\
\hline
Al    II     &         0.49 &     0.75 &     0.62 &        -5.94 &    -3.92 &    -3.15 &         1.00 &     0.90 &     0.87 \\
Al    III    &         1.05 &     0.96 &     0.83 &        -2.10 &    -1.92 &    -1.92 &         0.19 &     0.15 &     0.19 \\
Al    IV     &         1.29 &     0.98 &     0.87 &        -1.50 &    -1.54 &    -1.53 &         0.63 &     0.55 &     0.35 \\
Al    V      &         1.55 &     1.13 &     1.15 &        -0.75 &    -0.77 &    -0.90 &         0.55 &     0.52 &     0.30 \\
Al    VI     &         1.98 &     1.49 &     1.41 &        -0.03 &    -0.01 &    -0.31 &         0.58 &     0.58 &     0.63 \\
Al    VII    &         2.52 &     1.99 &     2.04 &         0.51 &     0.56 &     0.50 &         0.39 &     0.40 &     0.45 \\
Al    VIII   &         3.51 &     2.78 &     3.10 &         0.93 &     0.99 &     0.97 &         0.40 &     0.38 &     0.40 \\
Al    IX     &         5.75 &     4.15 &     4.92 &         1.29 &     1.36 &     1.30 &         0.43 &     0.45 &     0.42 \\
Al    X      &         9.95 &     7.09 &     8.24 &         1.59 &     1.65 &     1.60 &         0.28 &     0.29 &     0.27 \\
Al    XI     &        12.84 &    10.34 &    11.32 &         1.80 &     1.86 &     1.77 &         0.18 &     0.18 &     0.17 \\
Al    XII    &        20.23 &    15.74 &    15.73 &         2.10 &     2.16 &     2.07 &         0.59 &     0.58 &     0.59 \\
Al    XIII   &        68.71 &    61.99 &    41.85 &         2.43 &     2.46 &     2.41 &         0.48 &     0.47 &     0.48 \\
\hline
Si    II     &         0.58 &     0.88 &     0.76 &        -5.16 &    -2.69 &    -2.47 &         1.00 &     0.96 &     0.98 \\
Si    III    &         1.22 &     1.01 &     0.96 &        -1.68 &    -1.29 &    -1.24 &         0.60 &     0.14 &     0.33 \\
Si    IV     &         1.40 &     1.01 &     1.06 &        -1.17 &    -1.30 &    -1.07 &         0.38 &     0.22 &     0.30 \\
Si    V      &         1.57 &     1.08 &     1.23 &        -0.69 &    -0.89 &    -0.73 &         0.48 &     0.49 &     0.39 \\
Si    VI     &         1.82 &     1.32 &     1.41 &        -0.24 &    -0.30 &    -0.35 &         0.46 &     0.49 &     0.49 \\
Si    VII    &         2.26 &     1.77 &     1.88 &         0.30 &     0.33 &     0.33 &         0.52 &     0.53 &     0.56 \\
Si    VIII   &         3.51 &     2.78 &     2.68 &         0.93 &     1.02 &     0.84 &         0.59 &     0.61 &     0.41 \\
Si    IX     &         6.80 &     5.29 &     4.92 &         1.38 &     1.48 &     1.30 &         0.40 &     0.40 &     0.53 \\
Si    X      &        10.40 &     7.85 &     8.98 &         1.62 &     1.72 &     1.64 &         0.36 &     0.36 &     0.41 \\
Si    XI     &        13.22 &    11.15 &    13.80 &         1.83 &     1.90 &     1.94 &         0.22 &     0.22 &     0.24 \\
Si    XII    &        16.69 &    13.54 &    15.73 &         2.01 &     2.06 &     2.07 &         0.17 &     0.17 &     0.12 \\
Si    XIII   &        30.05 &    26.54 &    25.79 &         2.22 &     2.28 &     2.28 &         0.57 &     0.56 &     0.50 \\
Si    XIV    &        90.47 &    88.53 &    68.00 &         2.55 &     2.58 &     2.58 &         0.47 &     0.47 &     0.48 \\
\hline
S     II     &         0.72 &     0.56 &     0.41 &        -3.81 &    -7.66 &    -4.08 &         0.86 &     1.00 &     0.81 \\
S     III    &         1.20 &     0.96 &     0.75 &        -1.71 &    -1.87 &    -2.55 &         0.79 &     0.60 &     0.49 \\
S     IV     &         1.50 &     1.07 &     0.87 &        -0.87 &    -0.97 &    -1.53 &         0.49 &     0.54 &     0.78 \\
S     V      &         1.63 &     1.22 &     1.23 &        -0.57 &    -0.49 &    -0.73 &         0.17 &     0.14 &     0.50 \\
S     VI     &         1.72 &     1.30 &     1.36 &        -0.39 &    -0.35 &    -0.39 &         0.23 &     0.24 &     0.23 \\
S     VII    &         1.95 &     1.49 &     1.62 &        -0.06 &     0.02 &    -0.01 &         0.50 &     0.46 &     0.49 \\
S     VIII   &         2.57 &     2.12 &     2.29 &         0.54 &     0.63 &     0.67 &         0.57 &     0.59 &     0.61 \\
S     IX     &         4.18 &     3.42 &     4.36 &         1.08 &     1.19 &     1.22 &         0.48 &     0.46 &     0.47 \\
S     X      &         6.80 &     5.29 &     6.98 &         1.38 &     1.48 &     1.52 &         0.33 &     0.32 &     0.29 \\
S     XI     &        11.27 &     8.66 &    12.02 &         1.68 &     1.74 &     1.81 &         0.37 &     0.35 &     0.46 \\
S     XII    &        15.17 &    12.72 &    16.76 &         1.95 &     1.98 &     2.11 &         0.36 &     0.37 &     0.29 \\
S     XIII   &        21.97 &    17.57 &    22.65 &         2.13 &     2.18 &     2.24 &         0.22 &     0.23 &     0.20 \\
S     XIV    &        34.35 &    26.54 &    37.14 &         2.25 &     2.31 &     2.36 &         0.17 &     0.17 &     0.13 \\
S     XV     &        57.53 &    51.33 &    50.21 &         2.37 &     2.41 &     2.45 &         0.54 &     0.53 &     0.45 \\
S     XVI    &       134.35 &   122.81 &   105.30 &         2.73 &     2.77 &     2.79 &         0.47 &     0.46 &     0.47 \\
\hline
Ar    II     &         0.69 &     0.79 &     0.41 &        -4.14 &    -3.48 &    -4.08 &         0.63 &     0.93 &     0.78 \\
Ar    III    &         1.14 &     0.94 &     0.83 &        -1.86 &    -2.11 &    -1.87 &         0.92 &     0.45 &     0.92 \\
Ar    IV     &         1.50 &     0.98 &     0.96 &        -0.87 &    -1.59 &    -1.24 &         0.43 &     0.43 &     0.19 \\
Ar    V      &         1.72 &     1.05 &     1.06 &        -0.39 &    -1.04 &    -1.07 &         0.49 &     0.41 &     0.27 \\
Ar    VI     &         1.98 &     1.35 &     1.19 &        -0.03 &    -0.23 &    -0.81 &         0.36 &     0.72 &     0.49 \\
Ar    VII    &         2.15 &     1.77 &     1.36 &         0.18 &     0.35 &    -0.39 &         0.09 &     0.10 &     0.37 \\
Ar    VIII   &         2.26 &     1.82 &     1.62 &         0.30 &     0.41 &    -0.01 &         0.16 &     0.20 &     0.33 \\
Ar    IX     &         2.67 &     2.28 &     2.10 &         0.60 &     0.73 &     0.54 &         0.52 &     0.42 &     0.47 \\
Ar    X      &         4.03 &     3.23 &     3.46 &         1.05 &     1.13 &     1.05 &         0.46 &     0.46 &     0.46 \\
Ar    XI     &         7.20 &     5.29 &     6.98 &         1.41 &     1.50 &     1.52 &         0.41 &     0.40 &     0.45 \\
Ar    XII    &        12.07 &     9.50 &    12.68 &         1.74 &     1.78 &     1.86 &         0.36 &     0.37 &     0.34 \\
Ar    XIII   &        17.67 &    14.49 &    18.16 &         2.04 &     2.10 &     2.15 &         0.40 &     0.41 &     0.43 \\
Ar    XIV    &        30.05 &    26.54 &    30.94 &         2.22 &     2.29 &     2.32 &         0.30 &     0.30 &     0.25 \\
Ar    XV     &        45.56 &    38.05 &    41.85 &         2.31 &     2.36 &     2.41 &         0.18 &     0.18 &     0.18 \\
Ar    XVI    &        63.20 &    61.99 &    56.30 &         2.40 &     2.44 &     2.49 &         0.17 &     0.17 &     0.13 \\
Ar    XVII   &        90.47 &    88.53 &    74.03 &         2.55 &     2.59 &     2.62 &         0.52 &     0.51 &     0.43 \\
Ar    XVIII  &       253.22 &   254.93 &   164.07 &         2.88 &     2.89 &     2.91 &         0.46 &     0.47 &     0.47 \\
\hline
Ca    II     &         0.66 &     0.69 &     0.34 &        -4.38 &    -4.72 &    -4.46 &         0.93 &     0.77 &     0.93 \\
Ca    III    &         1.12 &     0.93 &     0.83 &        -1.92 &    -2.31 &    -2.00 &         0.73 &     0.63 &     0.78 \\
Ca    IV     &         1.44 &     0.98 &     0.92 &        -1.05 &    -1.54 &    -1.37 &         0.52 &     0.40 &     0.45 \\
Ca    V      &         1.63 &     1.04 &     1.06 &        -0.57 &    -1.08 &    -1.07 &         0.38 &     0.37 &     0.28 \\
Ca    VI     &         1.93 &     1.14 &     1.19 &        -0.09 &    -0.68 &    -0.86 &         0.45 &     0.39 &     0.28 \\
Ca    VII    &         2.36 &     1.35 &     1.26 &         0.39 &    -0.25 &    -0.65 &         0.46 &     0.39 &     0.34 \\
Ca    VIII   &         2.78 &     1.82 &     1.44 &         0.66 &     0.42 &    -0.26 &         0.22 &     0.68 &     0.45 \\
Ca    IX     &         3.05 &     2.55 &     1.84 &         0.78 &     0.90 &     0.29 &         0.07 &     0.10 &     0.47 \\
Ca    X      &         3.31 &     2.55 &     2.29 &         0.87 &     0.92 &     0.67 &         0.11 &     0.16 &     0.33 \\
Ca    XI     &         4.03 &     3.42 &     3.66 &         1.05 &     1.20 &     1.09 &         0.47 &     0.37 &     0.47 \\
Ca    XII    &         6.80 &     5.29 &     6.46 &         1.38 &     1.51 &     1.47 &         0.48 &     0.49 &     0.47 \\
Ca    XIII   &        12.07 &    10.34 &    11.32 &         1.74 &     1.85 &     1.77 &         0.41 &     0.43 &     0.37 \\
Ca    XIV    &        18.85 &    15.74 &    15.73 &         2.07 &     2.14 &     2.07 &         0.34 &     0.32 &     0.25 \\
Ca    XV     &        34.35 &    26.54 &    30.94 &         2.25 &     2.32 &     2.32 &         0.34 &     0.34 &     0.42 \\
Ca    XVI    &        57.53 &    51.33 &    50.21 &         2.37 &     2.41 &     2.45 &         0.27 &     0.27 &     0.28 \\
Ca    XVII   &        74.13 &    71.13 &    68.00 &         2.46 &     2.50 &     2.58 &         0.18 &     0.18 &     0.21 \\
Ca    XVIII  &        96.15 &    88.53 &    87.64 &         2.58 &     2.61 &     2.70 &         0.18 &     0.17 &     0.12 \\
Ca    XIX    &       134.35 &   122.81 &   105.30 &         2.73 &     2.76 &     2.79 &         0.50 &     0.50 &     0.39 \\
Ca    XX     &       424.40 &   370.90 &   291.44 &         2.97 &     2.96 &     3.00 &         0.46 &     0.46 &     0.46 \\
\hline
Fe    II     &         0.49 &     0.67 &     0.88 &        -5.94 &    -5.04 &    -1.49 &         1.00 &     0.66 &     0.52 \\
Fe    III    &         1.06 &     0.81 &     1.15 &        -2.07 &    -3.27 &    -0.90 &         0.74 &     0.55 &     0.51 \\
Fe    IV     &         1.31 &     0.92 &     1.29 &        -1.44 &    -2.35 &    -0.56 &         0.50 &     0.55 &     0.31 \\
Fe    V      &         1.43 &     0.98 &     1.41 &        -1.08 &    -1.60 &    -0.35 &         0.43 &     0.48 &     0.13 \\
Fe    VI     &         1.57 &     1.06 &     1.56 &        -0.69 &    -0.99 &    -0.09 &         0.43 &     0.46 &     0.51 \\
Fe    VII    &         1.86 &     1.27 &     2.04 &        -0.18 &    -0.42 &     0.50 &         0.51 &     0.46 &     0.50 \\
Fe    VIII   &         2.72 &     2.05 &     2.95 &         0.63 &     0.58 &     0.92 &         0.67 &     0.81 &     0.42 \\
Fe    IX     &         5.20 &     4.15 &     4.92 &         1.23 &     1.36 &     1.30 &         0.38 &     0.36 &     0.37 \\
Fe    X      &         8.08 &     6.41 &     6.98 &         1.47 &     1.62 &     1.52 &         0.35 &     0.37 &     0.36 \\
Fe    XI     &        10.84 &     9.50 &     9.75 &         1.65 &     1.82 &     1.69 &         0.25 &     0.29 &     0.27 \\
Fe    XII    &        12.07 &    11.94 &    12.02 &         1.74 &     1.95 &     1.81 &         0.16 &     0.21 &     0.18 \\
Fe    XIII   &        12.84 &    13.54 &    12.68 &         1.80 &     2.03 &     1.86 &         0.11 &     0.15 &     0.15 \\
Fe    XIV    &        13.22 &    14.49 &    13.80 &         1.83 &     2.08 &     1.94 &         0.07 &     0.09 &     0.11 \\
Fe    XV     &        14.08 &    14.49 &    14.35 &         1.89 &     2.12 &     1.98 &         0.04 &     0.05 &     0.06 \\
Fe    XVI    &        14.59 &    15.74 &    14.96 &         1.92 &     2.14 &     2.03 &         0.05 &     0.04 &     0.06 \\
Fe    XVII   &        16.69 &    17.57 &    16.76 &         2.01 &     2.21 &     2.11 &         0.39 &     0.22 &     0.30 \\
Fe    XVIII  &        24.06 &    26.54 &    22.65 &         2.16 &     2.30 &     2.24 &         0.39 &     0.35 &     0.38 \\
Fe    XIX    &        57.53 &    51.33 &    41.85 &         2.37 &     2.41 &     2.41 &         0.37 &     0.38 &     0.37 \\
Fe    XX     &        90.47 &    88.53 &    74.03 &         2.55 &     2.61 &     2.62 &         0.30 &     0.29 &     0.33 \\
Fe    XXI    &       124.42 &   122.81 &   105.30 &         2.70 &     2.73 &     2.79 &         0.26 &     0.27 &     0.30 \\
Fe    XXII   &       146.69 &   143.12 &   136.76 &         2.76 &     2.81 &     2.87 &         0.20 &     0.20 &     0.18 \\
Fe    XXIII  &       183.53 &   179.76 &   164.07 &         2.82 &     2.86 &     2.91 &         0.16 &     0.16 &     0.16 \\
Fe    XXIV   &       253.22 &   254.93 &   216.04 &         2.88 &     2.91 &     2.96 &         0.19 &     0.19 &     0.17 \\
Fe    XXV    &       496.94 &   510.00 &   378.39 &         3.00 &     3.00 &     3.04 &         0.49 &     0.50 &     0.45 \\
Fe    XXVI   &      1361.63 &  1297.50 &  1071.99 &         3.24 &     3.22 &     3.25 &         0.46 &     0.45 &     0.47 \\
\hline
Ni    II     &         0.57 &     0.68 &     0.60 &        -5.19 &    -4.94 &    -3.23 &         1.00 &     0.60 &     0.99 \\
Ni    III    &         1.20 &     0.78 &     0.85 &        -1.71 &    -3.59 &    -1.66 &         0.58 &     0.45 &     0.51 \\
Ni    IV     &         1.32 &     0.86 &     0.88 &        -1.41 &    -2.92 &    -1.49 &         0.28 &     0.40 &     0.16 \\
Ni    V      &         1.40 &     0.92 &     0.92 &        -1.17 &    -2.34 &    -1.37 &         0.45 &     0.47 &     0.14 \\
Ni    VI     &         1.53 &     0.97 &     0.96 &        -0.81 &    -1.65 &    -1.24 &         0.47 &     0.52 &     0.34 \\
Ni    VII    &         1.74 &     1.06 &     1.11 &        -0.36 &    -1.00 &    -0.98 &         0.40 &     0.46 &     0.44 \\
Ni    VIII   &         2.07 &     1.25 &     1.33 &         0.09 &    -0.45 &    -0.52 &         0.47 &     0.45 &     0.42 \\
Ni    IX     &         2.62 &     1.56 &     1.49 &         0.57 &     0.08 &    -0.18 &         0.39 &     0.49 &     0.41 \\
Ni    X      &         4.18 &     1.93 &     1.71 &         1.08 &     0.51 &     0.12 &         0.53 &     0.37 &     0.24 \\
Ni    XI     &         7.63 &     2.36 &     2.15 &         1.44 &     0.79 &     0.58 &         0.23 &     0.32 &     0.68 \\
Ni    XII    &        10.40 &     2.78 &     3.10 &         1.62 &     1.01 &     0.97 &         0.27 &     0.29 &     0.22 \\
Ni    XIII   &        13.63 &     3.42 &     3.87 &         1.86 &     1.20 &     1.13 &         0.33 &     0.28 &     0.31 \\
Ni    XIV    &        16.69 &     4.48 &     4.92 &         2.01 &     1.38 &     1.30 &         0.23 &     0.29 &     0.32 \\
Ni    XV     &        18.85 &     8.66 &     6.98 &         2.07 &     1.73 &     1.52 &         0.16 &     0.36 &     0.36 \\
Ni    XVI    &        21.97 &     7.09 &     8.24 &         2.13 &     1.67 &     1.60 &         0.07 &     0.32 &     0.32 \\
Ni    XVII   &        24.06 &    17.57 &    14.96 &         2.16 &     2.21 &     2.03 &         0.03 &     0.07 &     0.17 \\
Ni    XVIII  &        26.69 &    20.65 &    22.65 &         2.19 &     2.24 &     2.24 &         0.03 &     0.04 &     0.09 \\
Ni    XIX    &        34.35 &    38.05 &    25.79 &         2.25 &     2.34 &     2.28 &         0.32 &     0.25 &     0.36 \\
Ni    XX     &        51.65 &    51.33 &    41.85 &         2.34 &     2.41 &     2.41 &         0.37 &     0.35 &     0.26 \\
Ni    XXI    &        84.99 &    79.77 &    50.21 &         2.52 &     2.56 &     2.45 &         0.35 &     0.34 &     0.35 \\
Ni    XXII   &       124.42 &   122.81 &    74.03 &         2.70 &     2.73 &     2.62 &         0.27 &     0.27 &     0.28 \\
Ni    XXIII  &       162.44 &   143.12 &    95.65 &         2.79 &     2.82 &     2.74 &         0.24 &     0.25 &     0.27 \\
Ni    XXIV   &       212.96 &   179.76 &   113.92 &         2.85 &     2.87 &     2.83 &         0.19 &     0.19 &     0.22 \\
Ni    XXV    &       302.62 &   254.93 &   164.07 &         2.91 &     2.92 &     2.91 &         0.16 &     0.16 &     0.19 \\
Ni    XXVI   &       424.40 &   370.90 &   216.04 &         2.97 &     2.97 &     2.96 &         0.19 &     0.19 &     0.26 \\
Ni    XXVII  &       658.51 &   669.59 &   474.38 &         3.06 &     3.06 &     3.08 &         0.49 &     0.49 &     0.52 \\
Ni    XXVIII &      1669.70 &  1857.70 &  1495.96 &         3.30 &     3.30 &     3.34 &         0.46 &     0.45 &     0.45 \\

\end{longtable}
\end{longtab}

\end{document}